\documentclass[11pt]{article}
\pdfoutput=1
\usepackage{epsfig,sint,cite,amsfonts,amssymb,amsmath,bbold}
\usepackage[font=small,labelfont={sf,bf}]{caption}
\usepackage{subcaption}
\usepackage[T1]{fontenc}
\usepackage{tikz}
\usetikzlibrary{decorations.pathreplacing}
\usetikzlibrary{patterns} 
\usepackage[english]{babel}
\usepackage{simplewick}
\usepackage{dsfont}
\usepackage[utf8]{inputenc}
\usepackage{physics}
\usepackage{siunitx}
\usepackage{booktabs}
\usepackage{wrapfig}
\usepackage{slashed}

\newcommand{\be}{\begin{equation}}
\newcommand{\ee}{\end{equation}}
\newcommand{\bea}{\begin{eqnarray}}
\newcommand{\eea}{\end{eqnarray}}
\newcommand{\bes}{\begin{eqnarray}}
\newcommand{\ees}{\end{eqnarray}}
\newcommand{\ba}{\begin{array}}
\newcommand{\ea}{\end{array}}

\usetikzlibrary{decorations.markings}
\usetikzlibrary{svg.path}
\usetikzlibrary{shapes,fit}

\renewcommand{\i}{\mathrm{i}}

\begin{document}

\begin{titlepage}
\begin{flushright}
\hfill DESY 19-046\\
\end{flushright}

\vskip 0.5cm

\begin{center}

 {\Large\bf Frequency-splitting estimators of single-propagator traces\\[0.5ex]} 

\end{center}
\vskip 0.75 cm
\begin{center}
{\large  Leonardo Giusti and Tim Harris}
\vskip 0.25cm
Dipartimento di Fisica, Universit\`a di Milano--Bicocca,\\
and INFN, sezione di Milano--Bicocca,\\
Piazza della Scienza 3, I-20126 Milano, Italy\\
\vskip 1.0cm

{\large  Alessandro Nada and Stefan Schaefer}
\vskip 0.25cm
John von Neumann Institute for Computing (NIC),\\
DESY, Platanenallee 6, D-15738 Zeuthen, Germany\\
\vskip 0.75cm

{\bf Abstract}
\vskip 0.35ex
\end{center}

\noindent
Single-propagator traces are the most elementary fermion Wick contractions
which occur in numerical lattice QCD, and are usually computed by introducing
random-noise estimators to profit from volume averaging. The additional contribution
to the variance induced by the random noise is typically orders of magnitude larger
than the one due to the gauge field.
We propose a new family of stochastic estimators of single-propagator traces
built upon a frequency splitting combined with a hopping expansion of the quark
propagator, and test their efficiency in two-flavour QCD with pions as light as
$190$~MeV.
Depending on the fermion bilinear considered, the cost of computing these
diagrams is reduced by one to two orders of magnitude or more with respect to
standard random-noise estimators.
As two concrete examples of physics applications, we compute the disconnected
contributions to correlation functions of two vector currents in the isosinglet
$\omega$ channel and to the hadronic vacuum polarization relevant for the muon
anomalous magnetic moment. In both cases, estimators with variances dominated by
the gauge noise are computed with a modest numerical effort. Theory suggests large
gains for disconnected three and higher point correlation functions as well.
The frequency-splitting estimators and their split-even components are directly applicable
to the newly proposed multi-level integration in the presence of fermions.
\vfill
\eject

\end{titlepage}

\section{Introduction}
Disconnected fermion Wick contractions contribute to many physics processes at the forefront
of research in particle and nuclear physics: the hadronic contribution to the
muon anomalous magnetic moment, $K\rightarrow\pi\pi$ decays, nucleon form factors,
quantum electrodynamics and strong isospin-breaking contributions to hadronic matrix elements,
$\eta'$ propagator to name a few. When computed numerically in lattice Quantum Chromodynamics (QCD) and
if the distances between the disconnected pieces are large, their variances are dominated
by the vacuum contribution. The latter are well approximated by the product of variances of
the connected sub-diagrams the contractions are made of. The recently-proposed
multi-level Monte Carlo integration in the presence of fermions~\cite{Ce:2016idq,Ce:2016ajy} is
particularly appealing for computing disconnected contractions, since the various sub-diagrams
can be computed (essentially) independently from each other, thus making the scaling of the
statistical error with the cost of the simulation much more favorable with respect to the
standard Monte Carlo integration. 

The simplest examples of this kind are the disconnected Wick contractions of fermion bilinear
two-point correlation functions, where each single-propagator trace is usually computed by
introducing random-noise estimators~\cite{Bitar:1988bb,Dong:1993pk,Michael:1998sg}. As the action
of the auxiliary fields is already factorized, the multi-level
integration in the gauge field becomes highly profitable once the variance of each connected
sub-diagram is driven by its intrinsic gauge noise. The random-noise contribution, however,
is typically orders of magnitude larger than the one due to the gauge field, a fact which calls
for more efficient estimators in order to avoid the need of averaging over many random-noise fields
with large computational cost.

The aim of this paper is to fill this gap by introducing a new family of stochastic estimators of
single-propagator traces which combine the newly introduced split-even estimators with 
a frequency splitting and a hopping expansion of the quark propagator. We test their efficiency by
simulating two-flavour QCD with
pions as light as $190$~MeV. As a result, depending on the fermion bilinear considered, the cost
of computing single-propagator traces is reduced by one to two orders of magnitude or more
with respect to the
computational needs for standard random-noise estimators. The frequency-splitting estimators can be
straightforwardly implemented in any standard Monte Carlo computation of disconnected Wick
contractions, as well as directly combined with the newly proposed multi-level
integration in the presence of fermions.

In the next section we summarize basic facts about variances of generic disconnected
Wick contractions, while those of single-propagator traces are discussed in section~\ref{sec:SPTr}.
The following section is dedicated to introduce stochastic estimators of single-propagator traces
of heavy quarks based on a hopping expansion of the propagator, while
in section~\ref{sec:DeltaME} we introduce the split-even estimators for the difference of
two single-propagator traces also relevant for the muon anomalous magnetic moment.
The frequency-splitting estimators are introduced in section~\ref{sec:FSEs},
where also the outcomes of their numerical tests are reported.
In section~\ref{sec:twopts} we discuss the impact of these findings on two concrete examples of
physics applications: the disconnected contributions to the correlator of two electromagnetic currents
in the isospin limit relevant for the hadronic contribution to the muon anomalous magnetic moment, and 
the propagator of the $\omega$ vector meson. The paper ends with a short section of conclusions
and outlook, followed by some appendices where some useful notation and formulas are collected.

\section{Variances of disconnected Wick contractions\label{sec:disc}}
The connected correlation function of a generic disconnected Wick contraction, 
made of two sub-diagrams\footnote{Without loss of generality
we assume $W_i(x)$ to be real.} $W_0(0)$ and $W_1(x)$ centered at the origin and in
$x$ respectively, can be written as
\be
{\cal C}_{_{W_1 W_0}} =  \Big\langle \Big[W_1(x) - \langle W_1(x) \rangle \Big]
                      \Big[W_0(0) - \langle W_0(0) \rangle \Big] \Big\rangle\; , 
\ee
with its variance given by  
\be
\sigma^2_{_{{\cal C}_{W_1 W_0}}} =
\Big\langle
\Big[W_1(x) - \langle W_1(x) \rangle\Big]^2
\Big[W_0(0) - \langle W_0(0) \rangle\Big]^2
\Big\rangle - {\cal C}^2_{_{W_1 W_0}}\; . 
\ee
For large distances $|x|$,  
\be\label{eq:varprod}
\sigma^2_{_{{\cal C}_{W_1 W_0}}} = \sigma^2_{_{{\cal C}_{W_1}}} \cdot \sigma^2_{_{{\cal C}_{W_0}}}+\dots 
\ee
where
\be
\sigma^2_{_{{\cal C}_{W_0}}} = \Big\langle \Big[W_0(0) - \langle W_0(0) \rangle\Big]^2 \Big\rangle
\ee
and analogously for $\sigma^2_{_{{\cal C}_{W_1}}}$, and the dots stand for exponentially sub-leading effects. If
the gauge fields in the regions centered at the origin and in $x$ are updated independently
in the course of a multi-level Monte Carlo, e.g. Ref.~\cite{Ce:2016idq,Ce:2016ajy}, the
statistical error of each of the two sub-diagrams ${\cal C}_{W_0}$ and ${\cal C}_{W_1}$ decreases (essentially)
proportionally to the inverse of the square root of the cost of the simulation. The overall statistical
error on ${\cal C}_{W_1 W_0}$ thus scales with the inverse of the cost rather than with its square root. The above
argument can be iterated straightforwardly to multi-disconnected contractions.

Maybe the simplest example of this kind is a disconnected Wick contraction of the
correlator of two bilinear operators for which, following Eq.~(\ref{eq:varprod}),
the variance is well approximated by the product of variances of two
single-propagator trace estimators.

\section{Single-propagator traces\label{sec:SPTr}}
The single traces we are interested in are
\be\label{eq:tgamm}
t_{_{\Gamma,r}}(x) = - \frac{a_{_\Gamma}}{a^4} \tr \left[\Gamma D^{-1}_{m_r}(x,x) \right]\, ,
\ee
where $D_{m_r}$ is the massive Dirac operator with bare quark mass $m_r$ (for definiteness
we adopt the $O(a)$-improved Wilson-Dirac operator, see Appendix~\ref{app:Dw}),
$a$ is the lattice spacing, the factor
\be
a_{_\Gamma} = \left\{\begin{array}{cc}
1 & \Gamma=I, \gamma_5, \gamma_\mu\gamma_5,\sigma_{\mu\nu}\\ 
-i  & \hspace{-2.0cm}\Gamma=\gamma_\mu\\      
\end{array}\right. 
\ee
is chosen so that $t_{_{\Gamma,r}}(x)$ is real, and $\sigma_{\mu\nu}=\frac{i}{2}[\gamma_\mu,\gamma_\nu]$.
We are interested in the zero three-momentum field\footnote{Throughout this paper we focus on zero three-momentum fields
only. All techniques presented, however, are directly applicable to fields with non-zero three momentum.}
\be\label{eq:tgammP0}
\bar t_{_{\Gamma,r}}(x_0) = \frac{1}{L^3} \sum_{\bf x} a^3\, t_{_{\Gamma,r}}(x)\, ,
\ee
whose expectation value is
\be
s_{_{\Gamma,r}} = \langle \bar t_{_{\Gamma,r}}(x_0) \rangle  = a_{_\Gamma} \langle\bar\psi_r(x) \Gamma \psi_r(x) \rangle  
\ee
where $\psi_r$ is a quark flavour of mass $m_r$, and $L^3$ is the three-dimensional lattice volume. The variance of
$\bar t_{_{\Gamma,r}}(x_0)$, 
\be
\sigma^2_{\bar t_{_{\Gamma,r}}} = \langle \bar t^2_{_{\Gamma,r}}(x_0)\rangle -  \langle \bar t_{_{\Gamma,r}}(x_0)\rangle^2\; , 
\ee
can be written as
\be
\label{eq:GV}
\sigma^2_{\bar t{_{\Gamma,r}}} =  \frac{a^2_{_\Gamma}}{L^3} \sum_{\bf x} a^3 \langle
O_{_{\Gamma, rr}}(0,{\bf x})\, O_{_{\Gamma, r'r'}}(0)\rangle_c\, ,
\ee
where
\be
O_{_{\Gamma, rs}}(x) = \bar\psi_r(x) \Gamma \psi_s(x)\; ,
\ee
the subscript $c$ stands as usual for connected, and $\psi_{r'}$ is a second flavour\footnote{If not present in the theory, a
  valence quark $\psi_{r'}$ of mass $m_r$ can be added to it~\cite{Luscher:2004fu}.} of mass $m_{r'}=m_{r}$. The operator
product expansion would predict generically that
$\sigma^2_{\bar t{_{\Gamma,r}}}$ diverges as $a^{-3}$. There are exceptions, however, depending on the
symmetries preserved by the regularization and on the operator implemented\footnote{If the regularization
preserves the vector-flavour symmetry and its conserved current is adopted, for instance, the corresponding
variance vanishes in the infinite volume limit.}. Moreover $\sigma^2_{\bar t{_{\Gamma,r}}}$ vanishes in the
free-theory limit $g_0\rightarrow 0$, and the first non-zero contribution appears at $O(g_0^4)$ or
higher in perturbation theory.

\subsection{Random-noise estimator\label{sec:RNV}}
We introduce random auxiliary fields (random sources)~\cite{Bitar:1988bb,Michael:1998sg} defined so that all
their cumulants are null with the exception of the two-point functions which satisfy
\be\label{eq:random}
\langle \eta^a_\gamma (x) \{\eta^{b}_\delta(y)\}^* \rangle = \delta^{ab} \delta_{\gamma\delta} \delta_{x y}\; , 
\ee
where $a,b$ and $\gamma,\delta$ are colour and spin indices respectively, and $x,y$ are lattice coordinates.
By using Eq.~(\ref{eq:random}), it is straightforward to prove that a random-noise estimator
of $s_{_{\Gamma,r}}$ is given by
\be\label{eq:tau}
\tau_{_{\Gamma,r}}(x) = -\frac{1}{a^4 N_s} \sum_{i=1}^{N_s}
{\rm Re} \left[a_{_\Gamma} \eta^\dagger_i(x) \Gamma \{D^{-1}_{m_r}\eta_i\}(x)\right]\, ,  
\ee
where $\eta_i$ are $N_s$ independent sources (colour and spin indices omitted from now on).
The variance of the zero-momentum estimator 
\be\label{eq:tauP0}
\bar \tau_{_{\Gamma,r}}(x_0) = \frac{1}{L^3} \sum_{\bf x} a^3\, \tau_{_{\Gamma,r}}(x)\, , 
\ee
reads
\bea
\sigma^2_{\bar\tau_{_{\Gamma,r}}} \hspace{-0.25cm} =  
\sigma^2_{\bar t_{_{\Gamma,r}}}\hspace{-0.25cm} - \frac{1}{2 L^3 N_s}\left\{
a_{_\Gamma}^2 \sum_{\bf x} a^3 \langle O_{_{\Gamma, rr'}} (0,{\bf x}) O_{_{\Gamma, r'r}} (0) \rangle 
\hspace{-0.05cm} + \hspace{-0.05cm}\frac{1}{a}
\sum_x a^4 \langle P_{rr'} (x) P_{r'r} (0) \rangle\right\}\label{eq:tauV2},
\eea
where again $\psi_r$ and $\psi_{r'}$ are two degenerate flavours of mass $m_r$, and
to simplify the notation we have introduced the usual definition $P_{rs}=O_{_{\gamma_5, rs}}$
for the pseudoscalar density (no time-dilution is used since we are interested in the
estimator at all times). The random-noise contribution to the variances in
Eq.~(\ref{eq:tauV2}) diverges proportionally to $a^{-3}$ like the gauge one.
Both integrated correlators on the r.h.s. of Eq.~(\ref{eq:tauV2}), however, are
colour enhanced with respect to the gauge noise and they are of $O(1)$ in the
free theory, see Appendix \ref{app:free}.
The $\Gamma$-dependent contribution is indeed the flavour-connected counterpart of the
disconnected contraction appearing in Eq.~(\ref{eq:GV}). The $\Gamma$-independent
term $\langle PP \rangle$, which is also integrated over the time-coordinate,
diverges proportionally to $m_r^{-1}$ when $m_r \rightarrow 0$ due to the pion pole, giving
large contributions to the stochastic variances of all bilinears indistinctly. It is interesting
to notice that if we would not take the real part in Eq.~(\ref{eq:tau}), the variances would be larger
and $\Gamma$-independent since the $\langle O_{_{\Gamma, rr'}} O_{_{\Gamma, r'r}}\rangle$
contributions are dropped, and the prefactor $1/(2N_s)$ goes into $1/N_s$.

\begin{table*}
\small
\begin{center}
\setlength{\tabcolsep}{.10pc}
\begin{tabular}{@{\extracolsep{0.4cm}}ccccccccc}
\hline
id &$L/a$&$\kappa$&MDU&$N_{\rm cfg}$&$M_\pi$[MeV]&$M_\pi L$\\
\hline
E5  &$32$&$0.13625$ &$12800$& $100$ &$440$ &$4.7$ \\
F7  &$48$&$0.13638$ &$9600$&  $100(1200)$ &$268$ &$4.3$ \\
G8  &$64$&$0.136417$&$820$&   $25$ &$193$ &$4.1$ \\
\hline
\end{tabular}
\end{center}
\caption{\label{tab:ens} Overview of the ensembles and statistics used in this
study. We give the label, the spatial extent of the lattice, the hopping parameter
$\kappa$, the number of MDUs simulated after thermalization, the number of independent
configurations selected $N_{\rm cfg}$, the pion mass $M_\pi$, and the product $M_\pi L$.
For F7, $N_{\rm cfg}=100$ configurations have been used for estimating the variances while the
final results for the two-point functions have been obtained with $N_{\rm cfg}=1200$.}
\end{table*}
The random-noise contributions to the variances of the standard stochastic estimators in
Eq.~(\ref{eq:tau}) are thus expected to be much larger than the gauge-noise with
large ultraviolet and infrared divergent terms.

\subsection{Numerical tests}
To test the efficiency of the various stochastic trace estimators considered in this paper, we
have simulated QCD with two dynamical flavours discretized by the Wilson gluonic action and
the $O(a)$-improved Wilson--Dirac operator as defined in Appendix~\ref{app:Dw}. The details of
the ensembles of configurations considered, all generated by the CLS community
\cite{DelDebbio:2006cn,DelDebbio:2007pz,Fritzsch:2012wq}, are listed in Table~\ref{tab:ens}.
The bare coupling
constant is always fixed so that $\beta=6/g_0^2=5.3$, corresponding to a spacing of $\;a=0.065$\,fm. 
All lattices have a size of $2L \times L^3$, periodic boundary conditions for gluons, (anti-) periodic boundary
conditions in (time) space directions for fermions, and spatial dimensions always large enough so that $M_\pi L\geq4$.
The pion mass ranges from $190$ MeV to $440$ MeV. We have always skipped an enough number of 
molecular dynamics units (MDU) between two consecutive measurements so that gauge-field configurations
can be considered as independent in the statistical analysis, see Ref.~\cite{Fritzsch:2012wq,Engel:2014eea} and
references therein for more details.

\begin{figure}[t]
\begin{center}
\includegraphics[width=0.8\columnwidth]{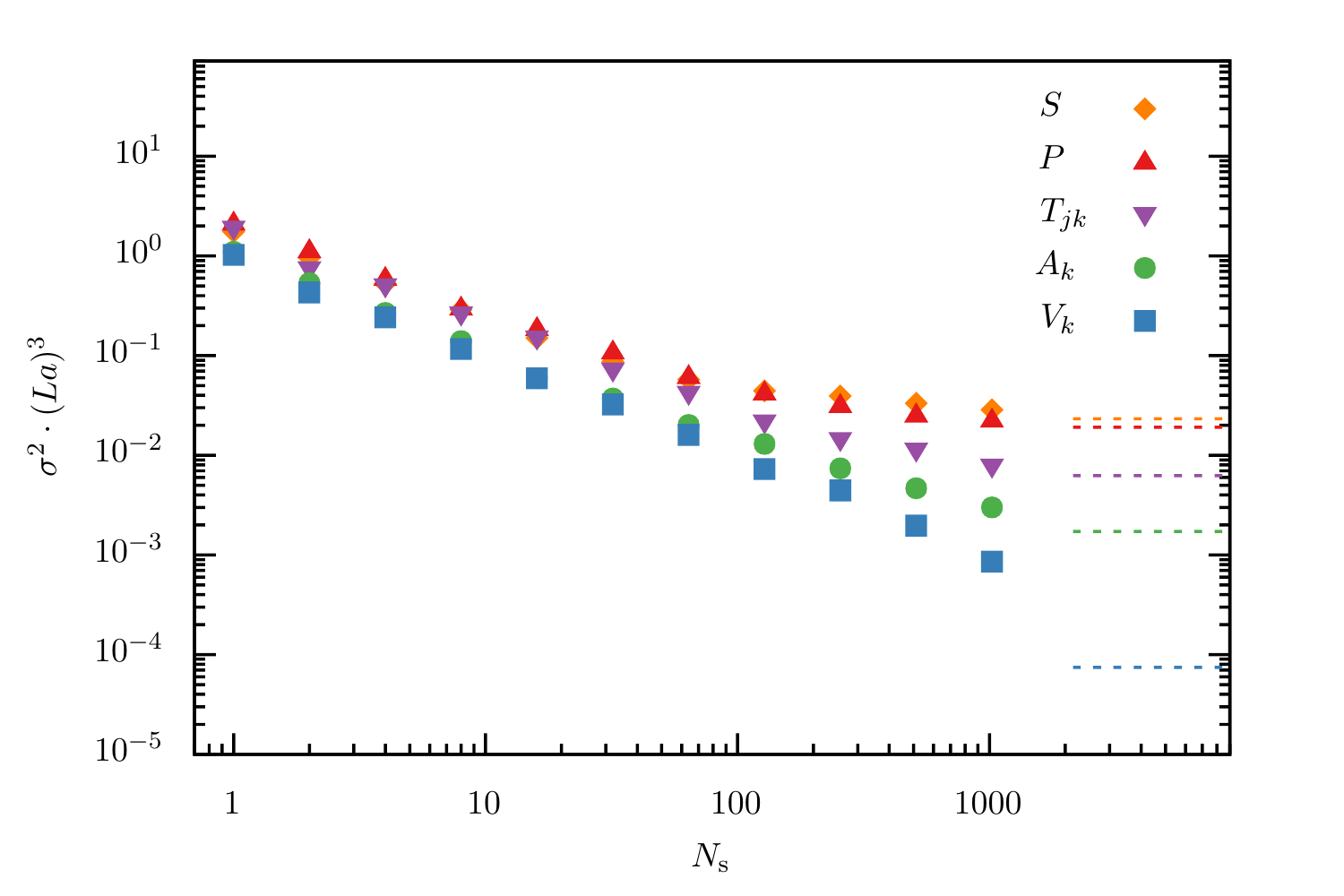}
\caption{Variances of the standard random noise estimators defined in Eq.~(\ref{eq:tau}) as a function
of the number of random sources $N_s$ for the ensemble F7. The symbols $S$, $P$, $T_{jk}$, $A_k$ and
$V_k$ stand for $\Gamma=I$, $\gamma_5$, $\sigma_{jk}$, $\gamma_k\gamma_5$ and
$\gamma_k$ respectively. The dashed lines indicate the gauge-noise contributions to the
variances computed in section \ref{sec:FSEs}. \label{Fig:std1}}
\end{center}
\end{figure}

The first primary observables that we have computed are the estimators
in Eq.~(\ref{eq:tauP0}) with Gaussian random noise. Their variances are shown in
Fig.~\ref{Fig:std1} as a function of the number of random-noise sources $N_s$  for
the ensemble F7. Data for the E5 and the G8 lattices show the same qualitative behaviour.
Variances go down linearly in $1/N_s$ until the random-noise contribution becomes
negligible, see Eq.~(\ref{eq:tauV2}), after which a plateau corresponds
to the gauge noise (dashed lines). The first clear message from the data is that
the random-noise contribution to the variances is comparable for the various bilinears,
as suggested by Eq.~(\ref{eq:tauV2}), and it is orders of magnitude larger than the
gauge noise. Moreover, the  latter can vary by orders of magnitude among the various
bilinears, see section \ref{sec:FSEs}, with the densities having the largest
gauge noise while the currents the smallest one.

\section{Hopping expansion of single-propagator traces\label{sec:HPEE}}
To investigate the contribution to trace variances from high-frequency modes
of the quark propagator, we first consider single-propagator traces of heavy quarks.
In this kinematic regime the hopping expansion (HPE) is known to lead to a
significant reduction of the random-noise contribution to trace
variances~\cite{Thron:1997iy,McNeile:2000xx,Bali:2009hu}. For the $O(a)$-improved
Wilson-Dirac operator, it is natural to exploit the
even-odd decomposition to generalize the hopping parameter expansion to
\be
D_m^{-1} = M_{2n,m} + D_m^{-1} H^{2n}_m\; ,   
\ee
where
\be\label{eq:HPEmain}
M_{2n,m} = \frac{1}{D_{\rm ee} + D_{\rm oo}} \sum_{k=0}^{2n-1} H^k_m\; , \qquad
H_m = -\left[D_{\rm eo} D_{\rm oo}^{-1} + D_{\rm oe} D_{\rm ee}^{-1}  \right]\; ,  
\ee
and the subscript $m$ has been omitted in the block matrices of the
even-odd decomposition of the Dirac operator, see Appendix \ref{app:Dw}
for further details.
\begin{figure}[t]
\begin{center}
\includegraphics[width=0.495\columnwidth]{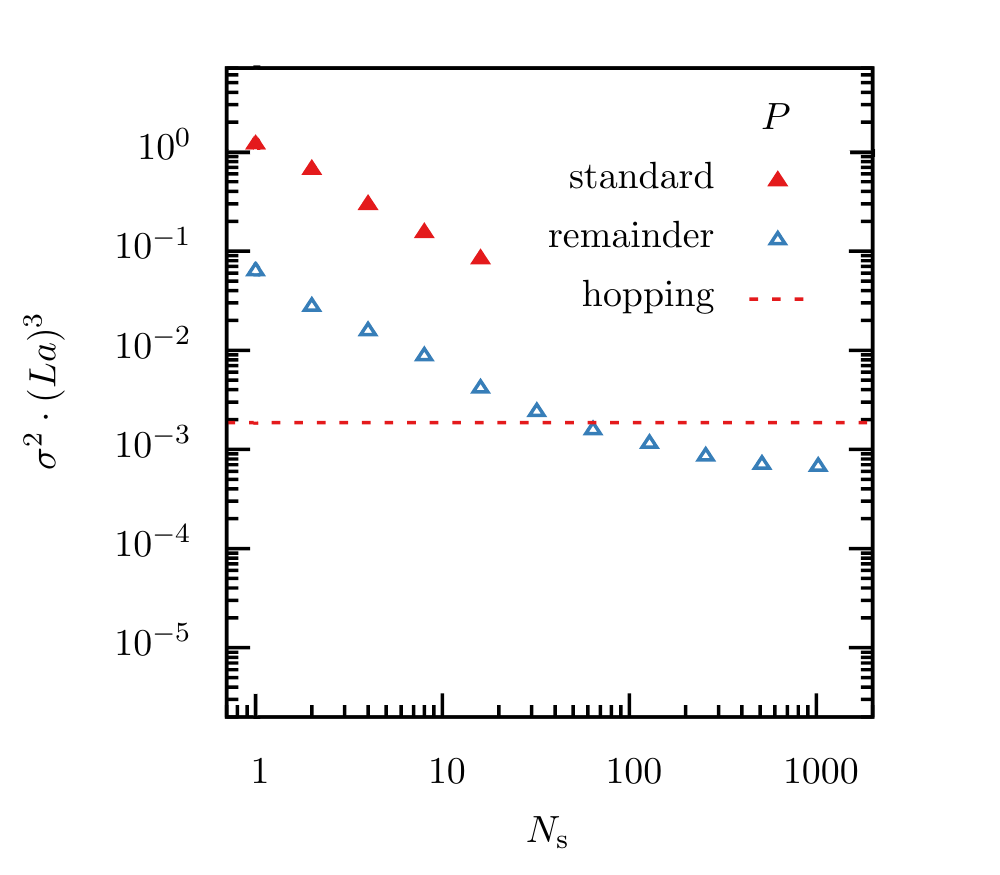}
\includegraphics[width=0.495\columnwidth]{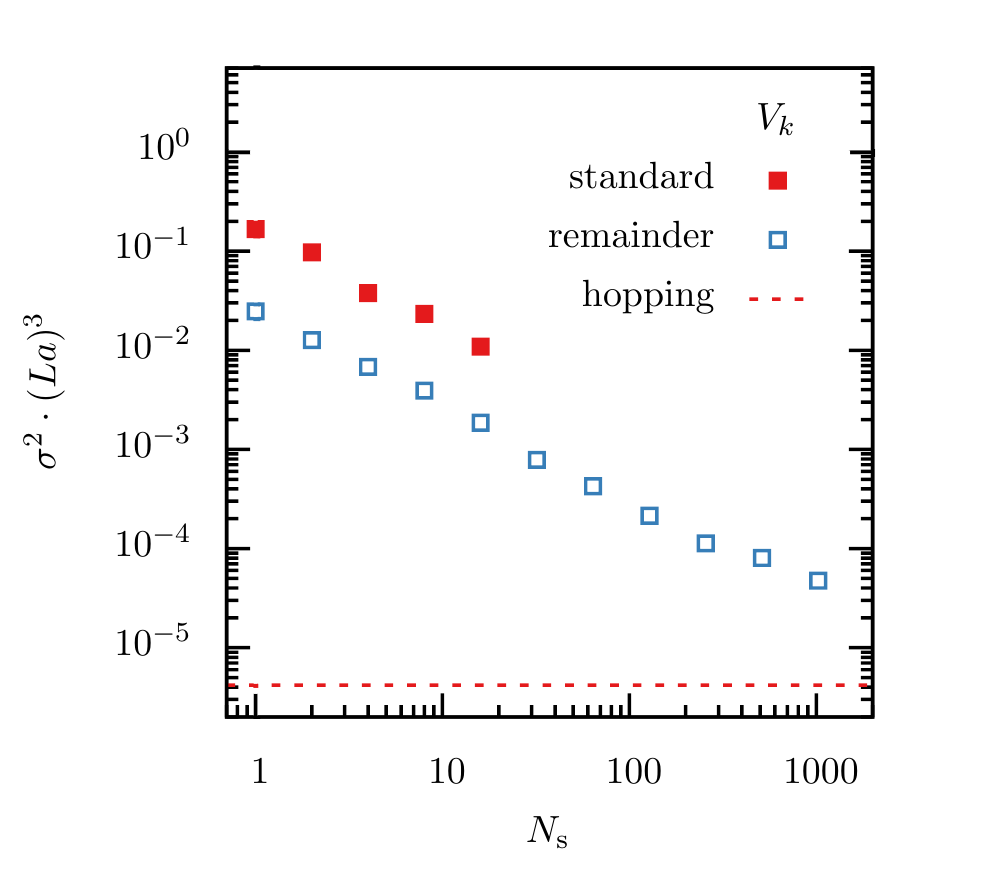}
\caption{Variances of the random-noise estimators $\tau^{R}_{_{\Gamma,r}}$ (remainder, open blue symbols)
for the pseudoscalar density (left) and the vector current (right) as a function of the number of random sources $N_s$ for
the ensemble F7, $n=2$, and a bare quark mass of $a m_{q,r}=0.3$. For comparison the variances 
of the standard random noise estimators (filled red symbols) and the gauge noise of
$\bar t^{M}_{_{\Gamma,r}}$ (hopping, dashed red lines) for the same mass are also shown.\label{Fig:hpe0.3}}
\end{center}
\end{figure}
The zero three-momentum single-propagator traces in Eq.~(\ref{eq:tgammP0}) can thus be
decomposed as
\be\label{eq:decompHPE}
\bar t_{_{\Gamma,r}}(x_0) = \bar t^{M}_{_{\Gamma,r}}(x_0) + \bar t^{R}_{_{\Gamma,r}}(x_0)\; , 
\ee
where
\be
\bar t^{M}_{_{\Gamma,r}}(x_0) =  - \frac{a_{_\Gamma}}{a L^3} \sum_{\bf x} \tr[\Gamma M_{2n,m_r}(x,x)]
\label{eq:thR}
\ee
collects the first $2 n$ contributions of the HPE while 
\be
\bar t^{R}_{_{\Gamma,r}}(x_0) =  - \frac{a_{_\Gamma}}{a L^3} \sum_{\bf x} \tr[\Gamma\{D_{m_r}^{-1} H^{2n}_{m_r} \}(x,x)]
\ee
is the remainder. Notice that convergence of the expansion is not required for Eq.~(\ref{eq:decompHPE})
to be valid. For small $n$, $\bar t^{M}_{_{\Gamma,r}}$ can be computed exactly and efficiently with $24\, n^4$ applications of
$M_{2n,m_r}$, see Appendix~\ref{app:mpls} for more details. The second contribution
$\bar t^{R}_{_{\Gamma,r}}$ can be replaced by the noisy estimator 
\be\label{eq:tauhR}
\bar \tau^{R}_{_{\Gamma,r}}(x_0) = - \frac{1}{a L^3 N_s} \sum_{\bf x} \sum_{i=1}^{N_s}
      {\rm Re}\left\{a_{_\Gamma}\big[\eta_{i}^{\dagger} H^{n}_{m_r}\big](x)\,
\Gamma\, \big[D_{m_r}^{-1}  H^{n}_{m_r} \eta_i\big](x) \right\} \; .
\ee
A rough idea of the variance reduction achieved by the HPE can be obtained in the free
lattice theory, see Appendix~\ref{app:free}. For a bare mass of $am=0.3$ and for $n=2$,
the stochastic variances of the remainder $\bar \tau^{R}_{_{\Gamma,r}}$ are between one
and two orders of magnitude smaller than those of the standard estimators
$\bar\tau_{_{\Gamma,r}}$. For $n=4$ a further reduction of approximately
4 to 8, depending on the bilinear, is obtained. If we had defined the estimator
of the remainder by applying $H^{2n}_{m_r}$ to one source only, the variance in the
free case would increase approximately by a factor 2 or so. The  ultraviolet filtering
of $H^n_{m_r}$ on both random sources is thus beneficial with respect to applying
$H^{2n}_{m_r}$ to one source only.

\subsection{Numerical tests}
We have computed the single-propagator trace estimators $\bar \tau_{_{\Gamma,r}}$,  
$\bar t^{M}_{_{\Gamma,r}}$ and $\bar\tau^{R}_{_{\Gamma,r}}$ for $n=2$ on all ensembles listed
in Table~\ref{tab:ens} for several valence quark masses. For F7 and for the subtracted bare quark
mass $a m_{q,r}=0.3$, the variances are shown in Fig.~\ref{Fig:hpe0.3} for the pseudoscalar
density and for a spatial component of the vector current respectively. Similar results
are obtained for other bilinears and/or for the E5 and the G8 lattices. The variances are in the same
ballpark as
the free-theory values. A clear picture emerges: the bulk of the random-noise contribution to
$\sigma^2_{\bar \tau{_{\Gamma,r}}}$ is due to $M_{2n,m_r}$ for all bilinears. Once the latter is subtracted
from the propagator and its contribution to $\bar t{_{_{\Gamma,r}}}$ is computed exactly, the random
noise is reduced by approximately one order of magnitude or more. Notice that
$\sigma^2_{\bar t^{M}_{_{\Gamma,r}}}$ is from 2 (pseudoscalar) up to 5 (vector) orders of magnitude smaller
than $\sigma^2_{\bar \tau{_{\Gamma,r}}}$ for $N_s=1$.
\begin{figure}[t]
\begin{center}
\includegraphics[width=0.495\columnwidth]{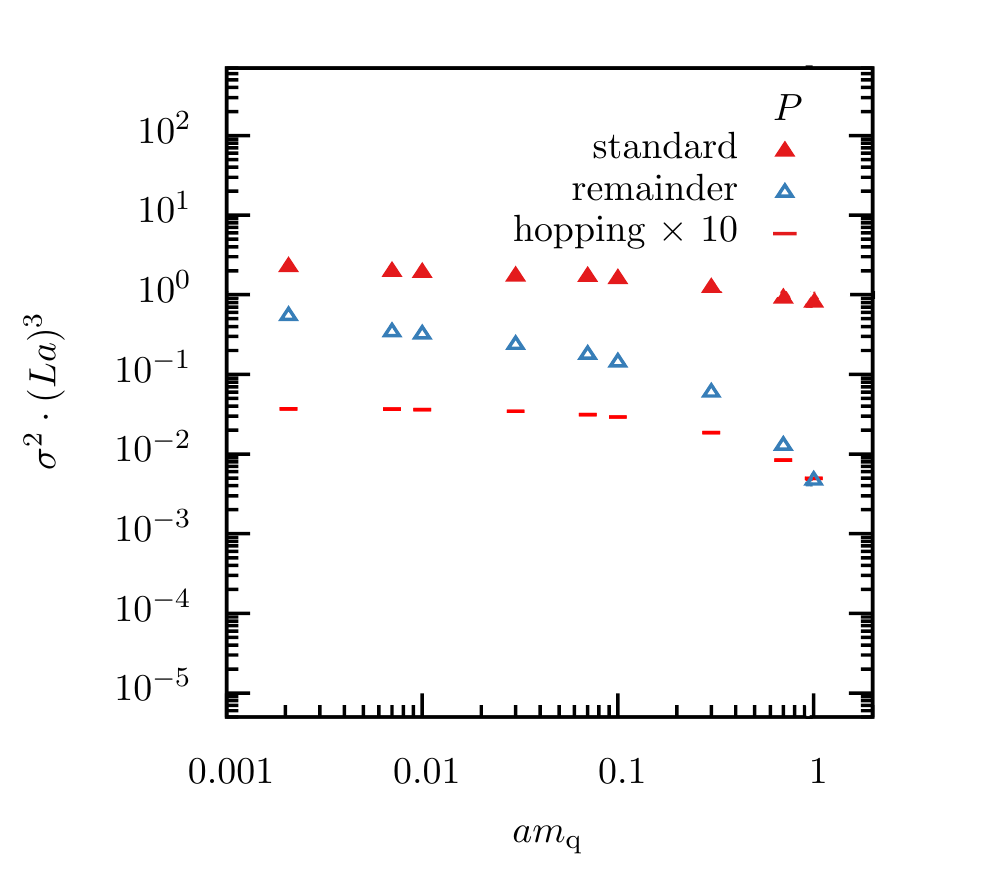}
\includegraphics[width=0.495\columnwidth]{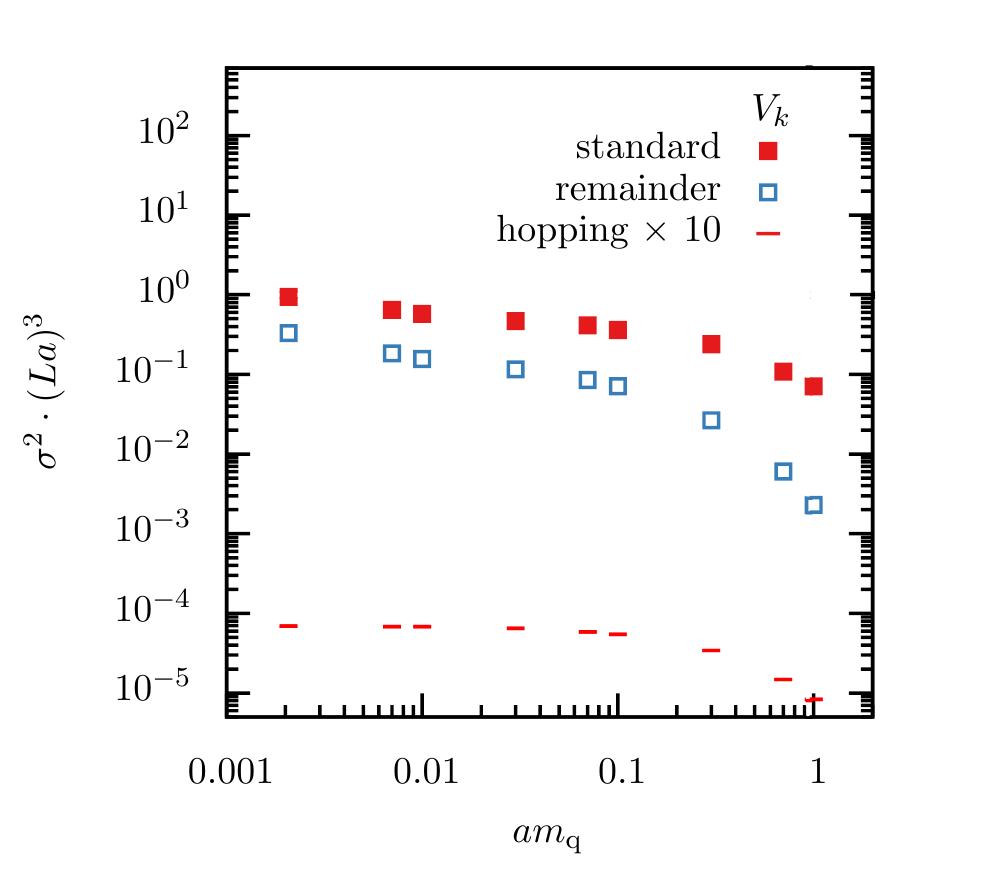}
\caption{Variances for $N_s=1$ of the random-noise estimators $\tau^{R}_{_{\Gamma,r}}$
  (remainder, open blue symbols) for the pseudoscalar density (left) and the vector current
  (right) as a function of the bare quark mass $a m_q$ for the ensemble F7 and $n=2$. For
  comparison the variances of the standard random noise estimators
  (filled red symbols) and the gauge noise of $\bar t^{M}_{_{\Gamma,r}}$ (hopping, line symbols)
  for the same mass are also shown.\label{Fig:hpeallm0}}
\end{center}
\end{figure}

In Fig.~\ref{Fig:hpeallm0}, $\sigma^2_{\bar\tau^{R}_{_{\Gamma,r}}}$ for $N_s=1$ and $\sigma^2_{\bar t^{M}_{_{\Gamma,r}}}$
multiplied by $10$ both for $n=2$ are shown as a function of the valence bare subtracted quark
mass $am_{q,r}$ for the pseudoscalar density and the spatial component of the vector current. As expected
the variance reduction due to the subtraction of $M_{2n,m_r}$ gets larger and larger at heavier quark
masses. In particular at $am_{q,r}=0.3$ the variance of the remainder is approximately one order of
magnitude smaller than at the sea quark mass value of $am_{q,r}=0.00207$. It is worth noting that even at
this light mass, the random-noise contribution to $\sigma^2_{\bar \tau{_{\Gamma,r}}}$ from $M_{2n,m_r}$
is still significant for all bilinears. The variance reduction due to HPE, however, is only a factor 2
or so. 

All in all data suggest that at heavy masses an efficient estimator of $s_{_{_{\Gamma,r}}}$ is obtained by
computing $\bar t^{M}_{_{\Gamma,r}}$ exactly and the remainder via the stochastic estimator
$\bar \tau^{R}_{_{\Gamma,r}}$. Which is the optimal order $n$ and how many random sources $N_s$ are required
for the remainder depend on the bilinear considered and on the final target observable of interest,
see section \ref{sec:FSEs}.

\section{Differences of single-propagator traces\label{sec:DeltaME}}
To analyse the contribution to trace variances from low-frequency modes
of the quark propagator, we consider the difference of two single-propagator traces
with different masses. It is worth noting, however, that often the difference itself is
a sub-diagram of the correlator of interest, e.g. the
disconnected contribution to the hadronic vacuum polarization from the up, down and
strange quarks in the exact isospin limit.
The estimator of the difference of two single-propagator traces reads 
\bea
t_{_{\Gamma,rs}}(x) & \equiv & t_{_{\Gamma,r}}(x) - t_{_{\Gamma,s}}(x)\nonumber\\[0.25cm] 
& = & - \frac{a_{_\Gamma}}{a^4}
\tr \left[\Gamma\{D^{-1}_{m_r}(x,x) - D^{-1}_{m_s}(x,x)\}\right]\nonumber\\[0.25cm]
& = & - \frac{a_{_\Gamma}}{a^4} (m_s-m_r)  \tr \left[\Gamma D^{-1}_{m_r} D^{-1}_{m_s}(x,x) \right]\, , 
\label{eq:tm1m2}
\eea
where $m_r\neq m_s$. Its expectation value can be written as 
\bea
s_{_{\Gamma,rs}} \equiv s_{_{\Gamma,r}} - s_{_{\Gamma,s}} & =  & a_{_\Gamma}\left\{
\langle O_{_{\Gamma, rr}} (x)\rangle -
\langle O_{_{\Gamma, ss}} (x)\rangle\right\} \nonumber\\[0.25cm]
 & = & a_{_\Gamma}\, (m_s-m_r) \sum_y a^4 \langle S_{rs}(y)\, O_{_{\Gamma, sr}} (x)) \rangle 
\eea
where, to simplify the notation, we have introduced the usual notation $S_{rs}=O_{_{I, rs}}$
for the scalar density. If we define the zero three-momentum field as
\be
\bar   t_{_{\Gamma,rs}}(x_0) = \frac{1}{L^3} \sum_{\bf x} a^3\, t_{_{\Gamma,rs}}(x)\; ,
\ee
its variance is given by
\be
\label{eq:GV2}
\sigma^2_{\bar t_{_{\Gamma,rs}}} = 
\frac{a^2_{_\Gamma}}{L^3} (m_s-m_r)^2\!\! \sum_{y_1,{\bf y_2},y_3}  a^{11} \langle S_{rs}(y_1)\,
O_{_{\Gamma, sr}} (0,{\bf y_2}) S_{s'r'}(y_3)\, O_{_{\Gamma, r's'}}(0) \rangle_c \, ,
\ee
where two extra valence fermions $\psi_{r'}$ and $\psi_{s'}$, with masses $m_{r'}=m_{r}$ and $m_{s'}=m_{s}$
respectively, are
introduced if not already present in the theory. This time the operator product expansion generically
predicts that $\sigma^2_{\bar t_{_{\Gamma,rs}}}$ diverges as $a^{-1}$, i.e. two powers less than in
Eq.~(\ref{eq:GV}) thanks to the presence of the squared-mass difference in the prefactor.
Analogously to section \ref{sec:SPTr}, there are exceptions depending on the symmetries
preserved by the regularization and on the discretization chosen for the operator, and
$\sigma^2_{\bar t_{_{\Gamma,rs}}}$ vanishes in the free-theory limit with the first non-zero
contribution appearing at $O(g_0^4)$ or higher in perturbation theory.

\subsection{Standard random-noise estimator}
Maybe the simplest random-noise estimator of $s_{_{\Gamma,rs}}$ is 
\be
\theta_{_{\Gamma,rs}}(x)
=  -\frac{(m_s-m_r)}{a^4 N_s} \sum_{i=1}^{N_s} {\rm Re} \left[ a_{_\Gamma}\eta^\dagger_i(x) \Gamma
  \{D^{-1}_{m_r}D^{-1}_{m_s}\eta_i\}(x)\right]\, ,\label{eq:theta12}
\ee
where the variance of its zero three-momentum counterpart 
\be\label{eq:theta13}
\bar \theta{_{{_\Gamma,rs}}}(x_0) =  \frac{1}{L^3} \sum_{\bf x} a^3\, \theta_{_{\Gamma,rs}}(x)
\ee
is
\bea
\sigma^2_{\bar\theta_{_{\Gamma,rs}}}\!\!\!\!\!\!\! & = &\!\!\! \sigma^2_{\bar t_{_{\Gamma,rs}}}\hspace{-0.25cm} -
\frac{(m_s-m_r)^2}{2 L^3 N_s}
\left\{
a_{_\Gamma}^2\hspace{-0.25cm}\sum_{y_1,{\bf y_2},y_3}\hspace{-0.25cm} a^{11} \langle S_{rs}(y_1) O_{_{\Gamma, ss'}} (0,{\bf y_2}) S_{s'r'}(y_3) O_{_{\Gamma, r'r}} (0) \rangle + 
\right.\nonumber\\
& & \left.\hspace{3.375cm}
\frac{1}{a}
\sum_{y_1,y_2,y_3} a^{12} \langle S_{rs}(y_1) P_{ss'} (y_2) S_{s'r'}(y_3) P_{r'r} (0) \rangle\right\}\, .
\label{eq:rnediff1}
\eea
Generically the random-noise contribution on the r.h.s of (\ref{eq:rnediff1}) diverges proportionally to
$a^{-1}$ like the gauge variance. The $\Gamma$-independent contribution $\langle S P S P \rangle$ is one
of the spectral sums introduced in Ref.~\cite{Giusti:2008vb}. It is integrated over one time-coordinate
more with respect to the first term, and it gives large contributions to the stochastic variances of all
bilinears indistinctly. If we would not take the real part in
Eq.~(\ref{eq:theta12}), the variances would be larger and $\Gamma$-independent since the
$\langle S O S O\rangle$ contributions are dropped, and the
prefactor $1/(2N_s)$ goes into $1/N_s$.

\subsection{Split-even random-noise estimator}
An alternative random-noise estimator of the difference of two traces is
\be\label{eq:bellaoe}
\tau_{_{\Gamma,rs}}(x) =  -\frac{(m_s-m_r)}{a^4 N_s} \sum_{i=1}^{N_s}
{\rm Re}\left[a_{_\Gamma} \{\eta^\dagger_i  D^{-1}_{m_r}\}(x)\, \Gamma\, \{D^{-1}_{m_s}\eta_i\}(x)\right] \; .
\ee
The corresponding zero three-momentum field is  
\be\label{eq:bellaoe2}
\bar\tau_{_{\Gamma,rs}}(x_0) = \frac{1}{L^3} \sum_{\bf x} a^3\, \tau_{_{\Gamma,rs}}(x)\; , 
\ee
and its variance reads
\bea
\sigma^2_{\bar\tau_{_{\Gamma,rs}}} \hspace{-0.75cm} & = & \hspace{-0.25cm}
\sigma^2_{\bar t_{_{\Gamma,rs}}}\hspace{-0.25cm} - \frac{a_{_\Gamma}^2 (m_s-m_r)^2}{2 L^3 N_s}
\sum_{y_1,{\bf y_2},y_3} a^{11} \Big\{
\big\langle  S_{rs}(y_1)\, O_{_{\Gamma, ss'}}(0,{\bf y_2})\,
             S_{s'r'}(y_3) \, O_{_{\Gamma, r'r}} (0) \big\rangle \nonumber\\[0.25cm]
& & 
\hspace{4.0cm} + \big\langle  P_{rr'}(y_1)\, O_{_{\Gamma, r's'}}(0,{\bf y_2})\,
             P_{s's}(y_3) \, O_{_{\Gamma, sr}} (0) \big\rangle 
             \Big\}
\label{eq:RNV7}
\eea          
where again two extra valence fermion fields $\psi_{r'}$ and $\psi_{s'}$ with masses
$m_{r'}=m_{r}$ and $m_{s}=m_{s'}$ are introduced if not already present in the theory. Also
this time the operator product expansion predicts generically that $\sigma^2_{\bar\tau_{_{\Gamma,rs}}}$
diverges as $a^{-1}$, but with respect to the standard random-noise estimator the (large)
$\Gamma$-independent spectral sum $\langle SPSP\rangle$ is absent. The first four-point correlation function
on the r.h.s. of Eq.~(\ref{eq:RNV7}) is the flavour-connected counterpart of the disconnected
contraction appearing in Eq.~(\ref{eq:GV2}), the second is analogous but with scalar densities
replaced by the pseudoscalar ones and two flavour indices exchanged. Both integrated
correlators on the r.h.s. of Eq.~(\ref{eq:RNV7}) are colour enhanced with respect to the
gauge noise and they are of $O(1)$ in the free theory, see Appendix \ref{app:free}.

The random-noise contributions to the variances of the split-even estimators in
Eq.~(\ref{eq:bellaoe2}) are thus expected to be significantly smaller than for the standard
estimators\footnote{The split-even is an estimator for all times at once, as well as the standard
estimator in Eq.~(\ref{eq:theta13}) we compare with. If time-dilution was used in (\ref{eq:theta13}),
the computation of the estimator for all times would have been singificantly more expensive.}
of differences of single-propagator traces. This is not surprising since in
this case both sources, $\eta_i$ and $\eta_i^\dagger$, are ultraviolet filtered
by a quark propagator and the variance has one integral less in the time-coordinate
analogously to the case of time-diluted sources\footnote{By the same argument, if a split-line
estimator localized in a given region of space is chosen, the sum over ${\bf y_2}$ in Eq.~(\ref{eq:RNV7})
is restricted to that region.}.
With respect to the gauge variance, however, the random-noise contribution is still
expected to be larger.

\begin{figure}[t]
\begin{center}
\includegraphics[width=0.45\columnwidth]{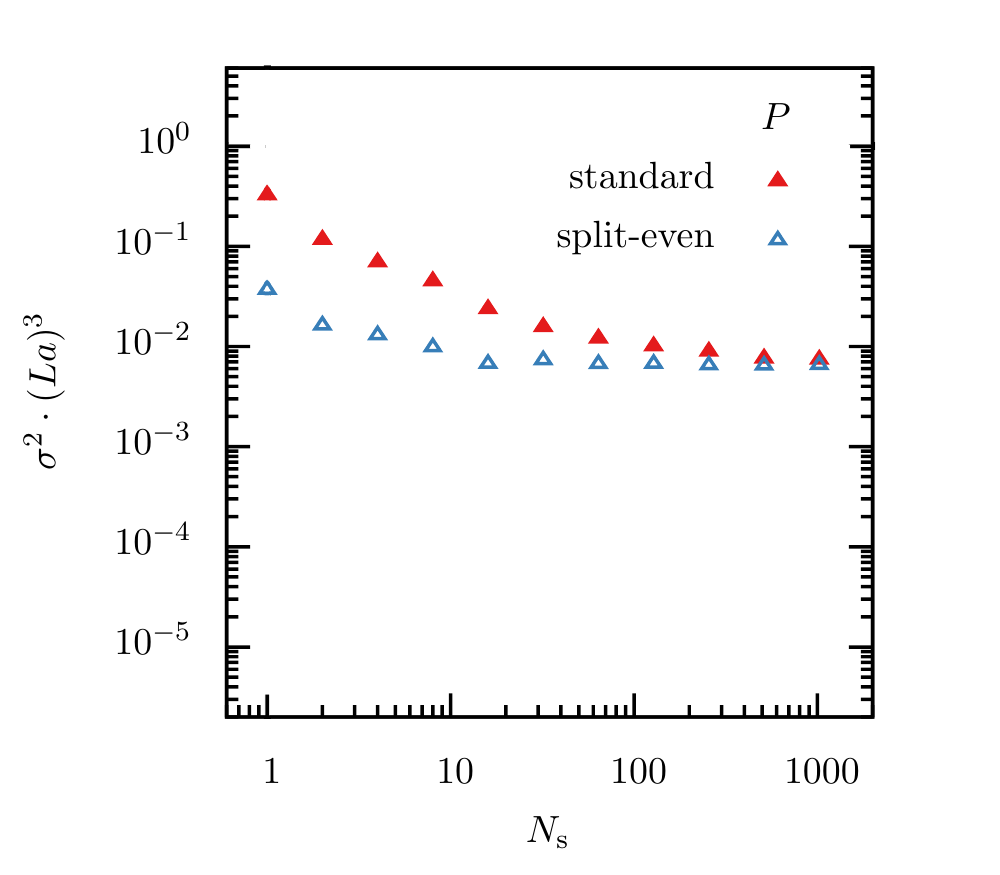}
\includegraphics[width=0.45\columnwidth]{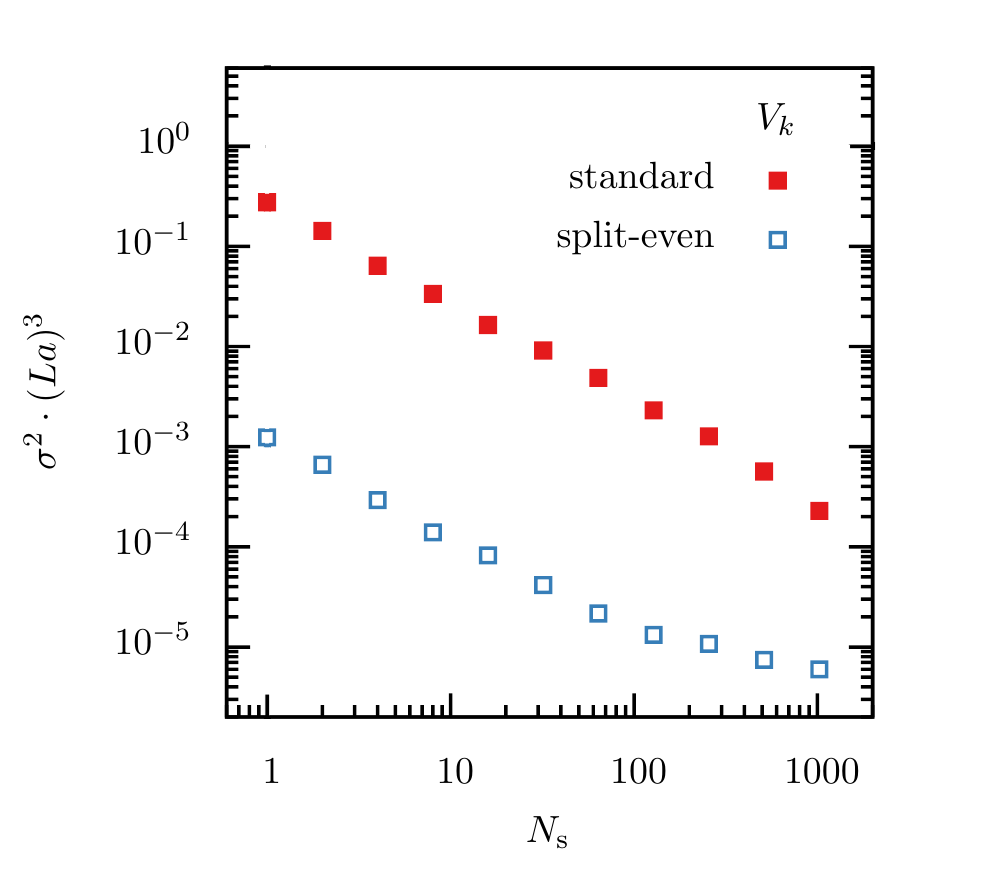}
\caption{Variances of the standard $\theta_{_{\Gamma,rs}}$ (filled red symbols) and the
  split-even $\tau_{_{\Gamma,rs}}$ (open blue symbols) estimators for differences of
  single-propagator traces for the pseudoscalar density (left) and the vector current
  (right) as a function of the number of random sources $N_s$ for the ensemble F7 and
  for the bare quark masses $am_{q,r}=0.00207$ and $am_{q,s}=0.0189$ corresponding to the
  sea and approximately the strange quark masses.\label{Fig:spln1}}
\end{center}
\end{figure}

\subsection{Numerical tests}
We have computed the two random-noise estimators in Eqs.~(\ref{eq:theta13})
and (\ref{eq:bellaoe2}) on all ensembles listed in Table~\ref{tab:ens} and for several
pairs of quark masses. For F7 and for the bare valence masses $am_{q,r}=0.00207$
and $am_{q,s}=0.0189$, corresponding to the sea and approximately the strange
quark masses~\cite{Fritzsch:2012wq,DellaMorte:2017dyu}, the variances
are shown in Fig.~\ref{Fig:spln1} for the pseudoscalar density and for one spatial
component of the vector current. Similar results are obtained for other bilinears
and/or other lattices. The variance of the standard estimators $\sigma^2_{\bar\theta_{_{\Gamma,rs}}}$
(red filled symbols) turns out to be essentially $\Gamma$-independent as suggested
by Eq.~(\ref{eq:rnediff1}), and it is dominated by the spectral sum
$\langle SPSP\rangle$. The split-even estimators $\bar\tau_{_{\Gamma,rs}}(x_0)$ have
much smaller variances\footnote{The so called one-end trick estimator used in the
context of twisted-mass discretization of QCD is a particular case of split-even
estimator, for which significant numerical gain has been observed
empirically~\cite{Boucaud:2008xu,Dinter:2012tt}.
The analysis of the variances presented here applies straightforwardly to this estimator too,
for which a Schwarz inequality between its variance and the one of the standard estimator
can also be derived.}.
The reduction factor ranges from approximately one order of
magnitude for the scalar and pseudoscalar densities up to around two orders of magnitude
or more for the axial and vector currents as well as for the tensor bilinear. The gauge noise
is still smaller than the random noise, but with the split-even estimator the
number $N_s$ of random sources needed to approach the gauge noise is moderate. It ranges
from a few for the pseudoscalar density up to $O(100)$ for the vector current. 

Overall, the data show that the split-even random-noise estimator is much more efficient 
than the standard one for computing differences of single-propagator traces, and
it allows one to approach the gauge noise for all bilinears with a moderate number
of noisy sources.

\section{Frequency-splitting of single-propagator traces\label{sec:FSEs}}
The results of the last two sections suggest to introduce a family of frequency-splitting
random-noise estimators of single-propagator zero three-momentum traces defined as
\be\label{eq:fsptheta}
\bar \tau^{\rm fs}_{_{\Gamma,r_1}}(x_0) = \bar t^{M}_{_{\Gamma,r_m}}(x_0) + \bar \tau^{R}_{_{\Gamma,r_m}}(x_0) +
\sum_{k=1}^{m-1} \bar \tau_{_{\Gamma,r_k r_{k+1}}}(x_0)\; , 
\ee
where $\bar t^{M}_{_{\Gamma,r_m}}$, $\bar \tau^{R}_{_{\Gamma,r_m}}$, and $\bar \tau_{_{\Gamma,r_k r_{k+1}}}$ are defined in
Eqs.~(\ref{eq:thR}), (\ref{eq:tauhR}), and (\ref{eq:bellaoe2}) respectively. The corresponding variances
are given by
\be
\sigma^2_{\bar\tau^{\rm fs}_{_{\Gamma,r_1}}}  = \sigma^2_{\bar t_{_{\Gamma,r_1}}} +
\{\sigma^2_{\bar \tau^{R}_{_{\Gamma,r_m}}}\hspace{-0.25cm} - \sigma^2_{\bar t^{R}_{_{\Gamma,r_m}}} \} +
\sum_{k=1}^{m-1}\{\sigma^2_{\bar \tau_{_{\Gamma,r_k r_{k+1}}}}\hspace{-0.5cm} - \sigma^2_{\bar t_{_{\Gamma,r_k r_{k+1}}}} \}\;  
\ee
where the various terms on the r.h.s are defined in sections \ref{sec:HPEE} and \ref{sec:DeltaME}.
At high momenta (heavy masses) the contribution from $\bar t^{M}_{_{\Gamma,r_m}}$, responsible for the bulk
of the variance of the standard random-noise estimator, is computed exactly with a limited number of probing vectors, and only
the remainder $\bar \tau^{R}_{_{\Gamma,r_m}}$ is estimated by a random-noise estimator. The low-frequency contributions
$\tau_{_{\Gamma,r_k r_{k+1}}}$ can then be estimated by the very efficient split-even estimator. It is rather clear that splitting the
single-propagator traces in several parts whose contributions come from different frequency regions is beneficial. It allows
us to design a customized estimator for each contribution which profits from its own peculiarities. An important ingredient in
this analysis is the fact that solvers invert the Dirac operator with heavier quark masses at a lower numerical cost.

\subsection{Numerical tests}
The best choice of the number of mass differences, the values of the masses, and the order of the HPE for defining the
frequency-splitting estimators in Eq.~(\ref{eq:fsptheta}) depends on many factors: the bilinear of interest,
the target mass, the solver chosen for inverting the Dirac operators and its particular implementation, etc.
It is not the aim of this paper to optimize with respect to all these factors\footnote{Computing variances for such an
optimization is cheap because it requires a few sources only.} but, provided a reasonable choice
is made, our goal is to give a numerical proof that the frequency-splitting estimators are efficient and allow to reduce
significantly the numerical cost for computing single-propagator traces. To this aim we have implemented two
such estimators:
\begin{figure}[t]
\begin{center}
\includegraphics[width=0.45\columnwidth]{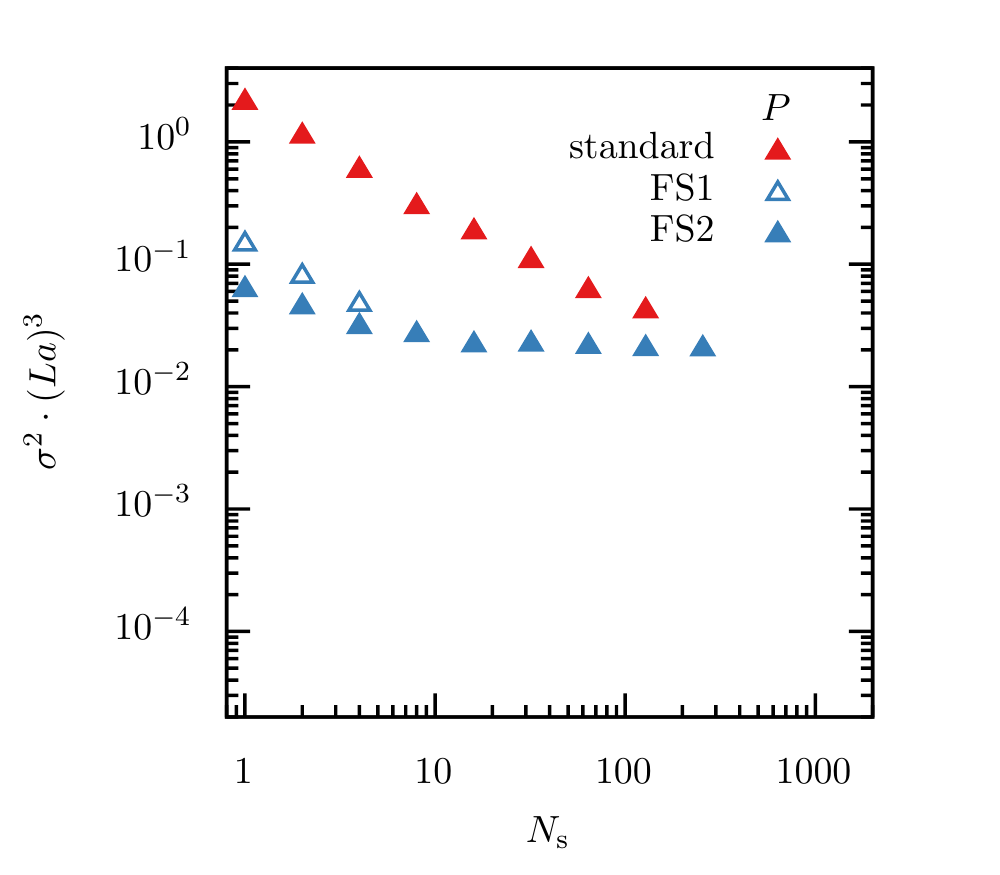}
\includegraphics[width=0.45\columnwidth]{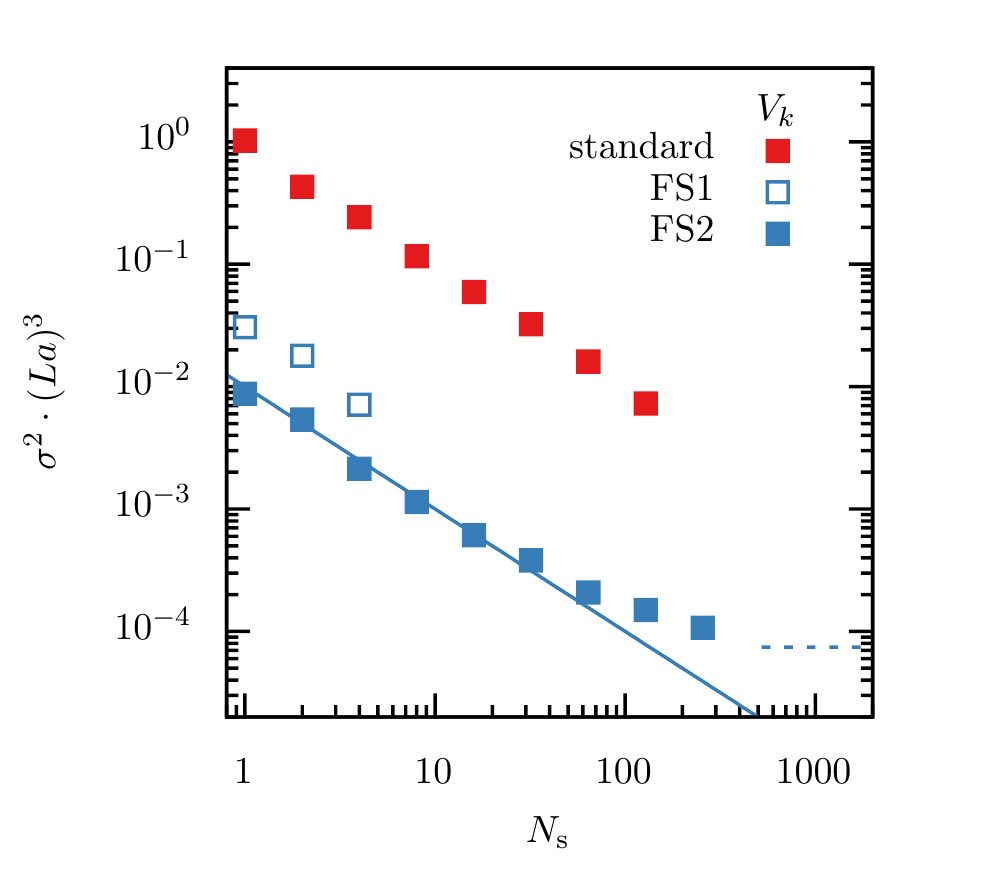}
\caption{Variances of the two frequency-splitting estimators FS1 and FS2 (see
  the main text for explanations) for the pseudoscalar density (left) and the
  vector current (right) for the ensemble F7 and for the target mass
  $am_q=0.00207$ corresponding to the sea quark mass. On the horizontal axis,
  $N_s$ is the number of times the frequency-splitting estimator is evaluated and
  averaged over for each gauge configuration. For comparison the variances of the
  standard random-noise estimators (filled red symbols) are also shown,  where in this
  case $N_s$ coincides with the number of random sources processed and averaged over
  for each gauge configuration. In the right plot, the continuum line represents the
  linear term of a linear fit in $1/N_s$ of the points, while the dashed line is the constant
  term corresponding to the gauge noise.
\label{Fig:FSEPV}}
\end{center}
\end{figure}
\begin{itemize}
\item  FS1 is the simplest frequency-splitting estimator with one mass
  difference only. The masses are $a m_q=0.00207$ and $0.1$, $\bar \tau_{_{\Gamma,r_0 r_{1}}}$ and
  $\bar \tau^{R}_{_{\Gamma,r_1}}$ are defined with $N_s=1$ and $4$ respectively. For the lattice F7, inverting the Dirac
  operator at the heavier mass costs approximately $1/3$ than at the target lighter mass. Each evaluation
  of this estimator therefore costs approximately $2.5$ times more than processing one random source for
  the standard estimator\footnote{We do not include the preparatory cost for computing $\bar t^{M}_{_{\Gamma,r_n}}$
  since it becomes quickly negligible after few evaluations of the random-noise components of the estimator.}. 
\item  FS2 is defined by 4 mass splittings corresponding to the masses $a m_q=0.00207$, $0.02$, $0.06$, $0.15$, $0.3$, and
  the corresponding random-noise estimators are defined with $N_s=1$, $1$, $2$, $3$, and $10$ random sources respectively.
  For the lattice F7, the cost of inverting the Dirac operator for the second up to the fifth mass is
  approximately $1/2$, $1/3$, $1/4$, and $1/6$ 
  with respect to the lightest quark mass respectively. Each application of this estimator thus costs approximately
  $6.5$ times with respect to processing one random source for the standard estimator.
\end{itemize}
In both cases the solver used is the generalized conjugate residual (GCR) algorithm preconditioned by 
a Schwarz alternating procedure (SAP) and local deflation as implemented in openQCD-1.6 \cite{openQCD1.6}.

In Fig.~\ref{Fig:FSEPV} we show the variances of FS1, FS2 and of the standard estimator as a function of $N_s$, the number 
of evaluations of each of them per gauge configuration. Similar plots are obtained for the other bilinears and the
other two lattices. A clear message emerges: a large gain is obtained for both frequency-splitting estimators  with mild
differences in efficiency between them. The FS1 is slightly better for the scalar and pseudoscalar densities, while FS2
is more efficient for the vector, axial-vector and tensor bilinears. In particular, the variance of FS1 is approximately
20 and 15 times smaller
than the one of the standard estimators for the scalar and pseudoscalar densities respectively. Taking into account that
one application of FS1 costs approximately 2.5 more, the gain in computation cost is 8 and 6 for the scalar and
pseudoscalar\footnote{If we had used $U(1)$ sources instead of Gaussian ones, the standard estimator for
the pseudoscalar density would have a variance smaller by approximately a factor $3$ on this lattice. We prefer to use
Gaussian sources for all bilinears, however, for which the theoretical analysis is simpler.} densities respectively.
For the vector and the axial-vector, the variance of FS2 is approximately 2 orders of magnitude
smaller than the one of the standard estimators. As the FS2 is 6.5 times more expensive, the gain in computational
cost is approximately a factor 15. For the tensor the factor gained reaches approximately 20.

It is worth noting that for the scalar, pseudoscalar and the tensor bilinears just one or a few evaluations of the
frequency-splitting estimators are needed for the variance to be comparable to the gauge noise. For the axial-vector
and vector currents $O(10)$ and $O(100)$ evaluations of the FS2 estimators are required to reach the same goal. As a
result, in all cases the gauge noise is reached with a limited and affordable number of evaluations of the
frequency-splitting estimators. If necessary the frequency-splitting estimator can be easily combined with
low-mode averaging \cite{Giusti:2004yp,DeGrand:2004qw} and its variants \cite{Blum:2012uh}.

\section{Numerical tests for two-point functions\label{sec:twopts}}
In this section we discuss the numerical results for two representative examples
of disconnected contributions to two-point functions, which are the simplest
correlation functions with a non-trivial time dependence composed only of
single-propagator traces. We use the estimators proposed in sections~\ref{sec:DeltaME}
and~\ref{sec:FSEs} to confirm the expected improvement over the standard estimator, and
check the factorization formula for the variance given in section~\ref{sec:disc}.

\subsection{Split-even estimator for electromagnetic current}
As alluded to in section~\ref{sec:DeltaME}, an important application of the
split-even estimator is the determination of the disconnected contribution to the
correlation function of two electromagnetic currents with three light flavours.
In the isospin limit, this gives rise to a difference of single-propagator
traces as in Eq.~(\ref{eq:tm1m2}) with $r$ and $s$ corresponding to the up/down
and strange quark flavours respectively. In particular the correlator
\begin{equation}
  C^{rs}_{VV}(x_0) = - \frac{L^3}{3 L_0}\sum_{k=1}^{3}\sum_{y_0} a\,
    \langle\bar t_{\gamma_k,rs}(x_0+y_0)\, \bar t_{\gamma_k,rs}(y_0)\rangle\, 
  \label{eq:twopt_em}
\end{equation}
determines the light disconnected contribution, via the time-momentum representation~\cite{Bernecker:2011gh},
of the leading-order hadronic vacuum polarization, once each current is renormalized by
$Z_V=0.74636(70)$ \cite{DallaBrida:2018tpn} and the correct
electric charge factor of $1/9$ is included.

\begin{figure}[t]
\begin{center}
\includegraphics[width=0.45\columnwidth]{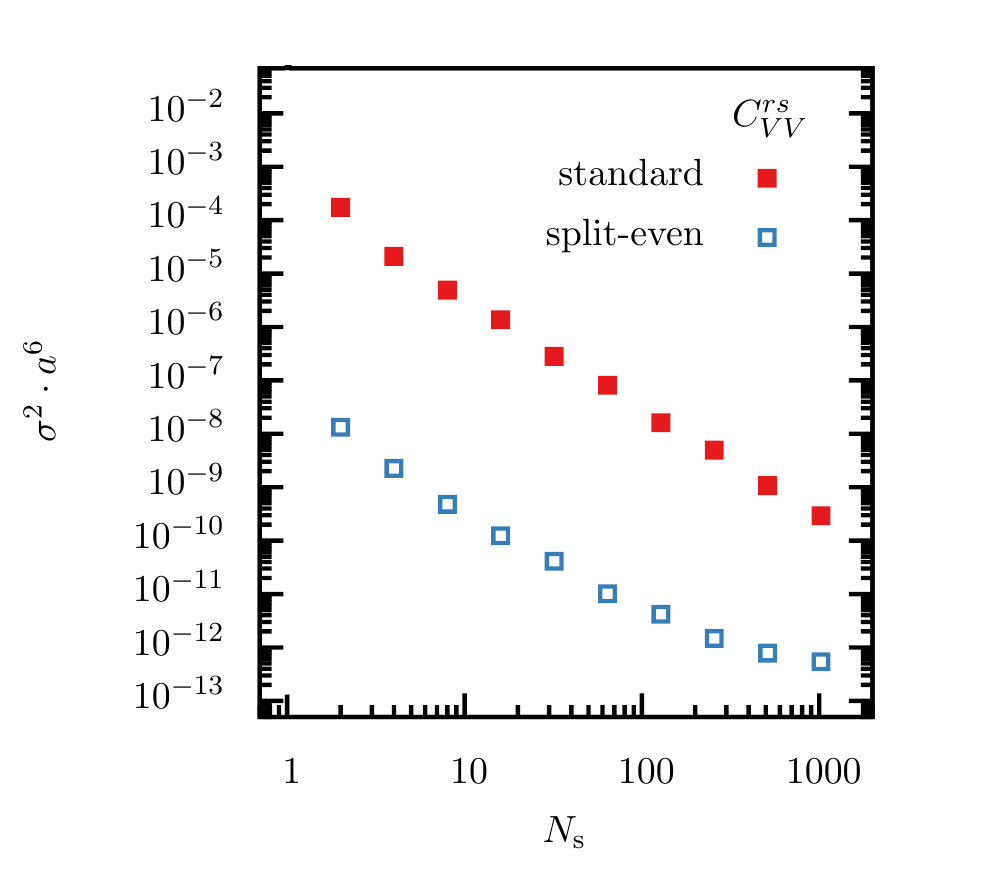}
\includegraphics[width=0.45\columnwidth]{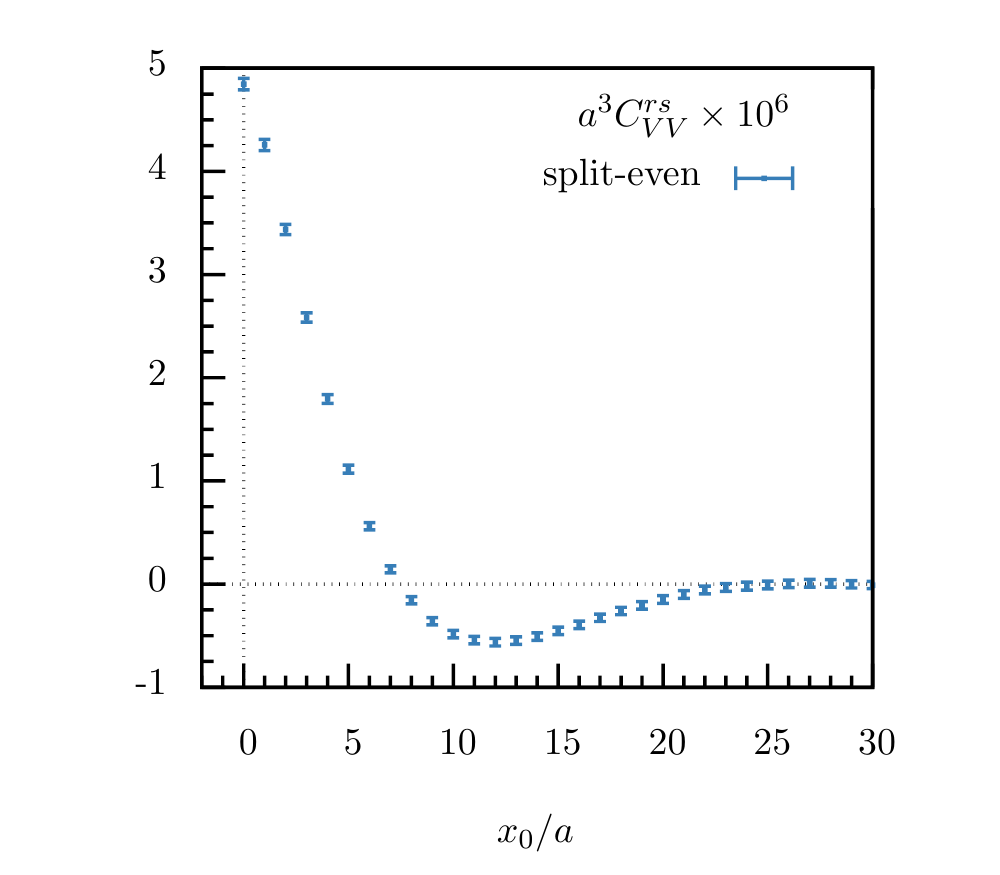}
\caption{Left: variance of the disconnected contribution in 
  Eq.~\eqref{eq:twopt_em} with $x_0/a=10$ using the standard (red filled
  squares) and split-even estimator (blue open squares).
  The stochastic noise of the split-even estimator is comparable with the gauge
  noise after $N_s\sim256$.
  Right: the disconnected contribution using the split-even estimator
  from $N_\mathrm{cfg}=1200$ gauge configurations.}
\label{Fig:twopt_em} 
\end{center}
\end{figure}

In the left-hand panel of Fig.~\ref{Fig:twopt_em}, we show the variance of
this correlation function for $x_0/a=10$ computed by using
the standard (red filled squares) and split-even estimators (blue open
squares) in Eqs.~\eqref{eq:theta13} and~\eqref{eq:bellaoe2} respectively.
A reduction of the variance of up to four orders of magnitude is obtained with
the split-even estimator (two orders of magnitude in the cost), which starts to
be comparable to the gauge noise for
$N_s\sim 256$. As expected, the variance is practically constant in $x_0$ and
well-described by the factorization formula in Eq.~\eqref{eq:varprod} when the
averaging over time and the polarizations of the current are taken into account.

In the right-hand panel of Fig.~\ref{Fig:twopt_em} our best estimate of the
correlation function using the split-even estimator is shown using an increased number of
gauge configurations, with respect to those used for estimating the variances, of
$N_\mathrm{cfg}=1200$. This in turn corresponds to a relative statistical precision of approximately $10\%$
to the disconnected light-quark part of the muon anomalous magnetic moment coming
from contributions to the integral up to time-distances of $1.5$~fm. If
the integral is computed up to $3.0$~fm or so, the relative statistical error
grows up to $70\%$, calling for the multi-level integration to determine the contribution
from the long distance part of the integrand. To properly
renormalize the correlator each current has to be multiplied by the factor
$Z_V$ which brings a negligible error with respect to the
statistical error of the bare correlator\footnote{Improving the vector current goes beyond
the scope of this paper. All formulas, however, can be found in
Ref.~\cite{Gerardin:2018kpy,Bhattacharya:2005rb}.}.

\subsection{Frequency-splitting estimator for isoscalar vector currents}
\begin{figure}[t]
\begin{center}
\includegraphics[width=0.45\columnwidth]{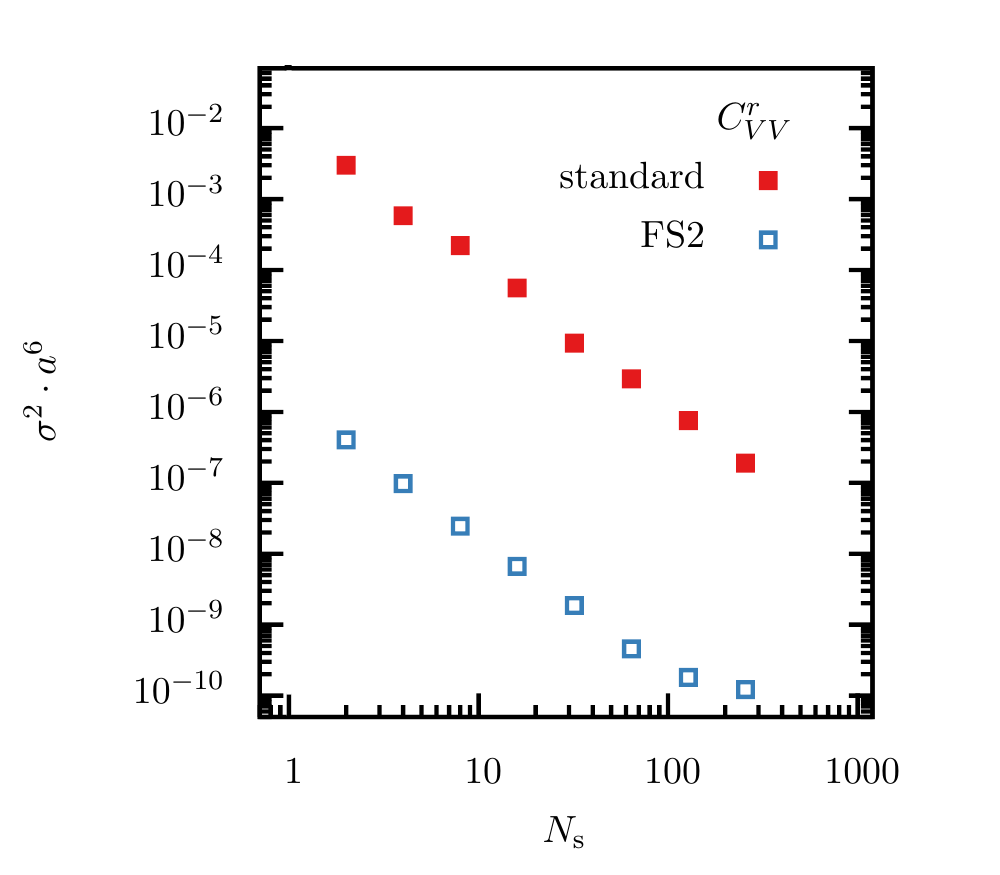}
\includegraphics[width=0.45\columnwidth]{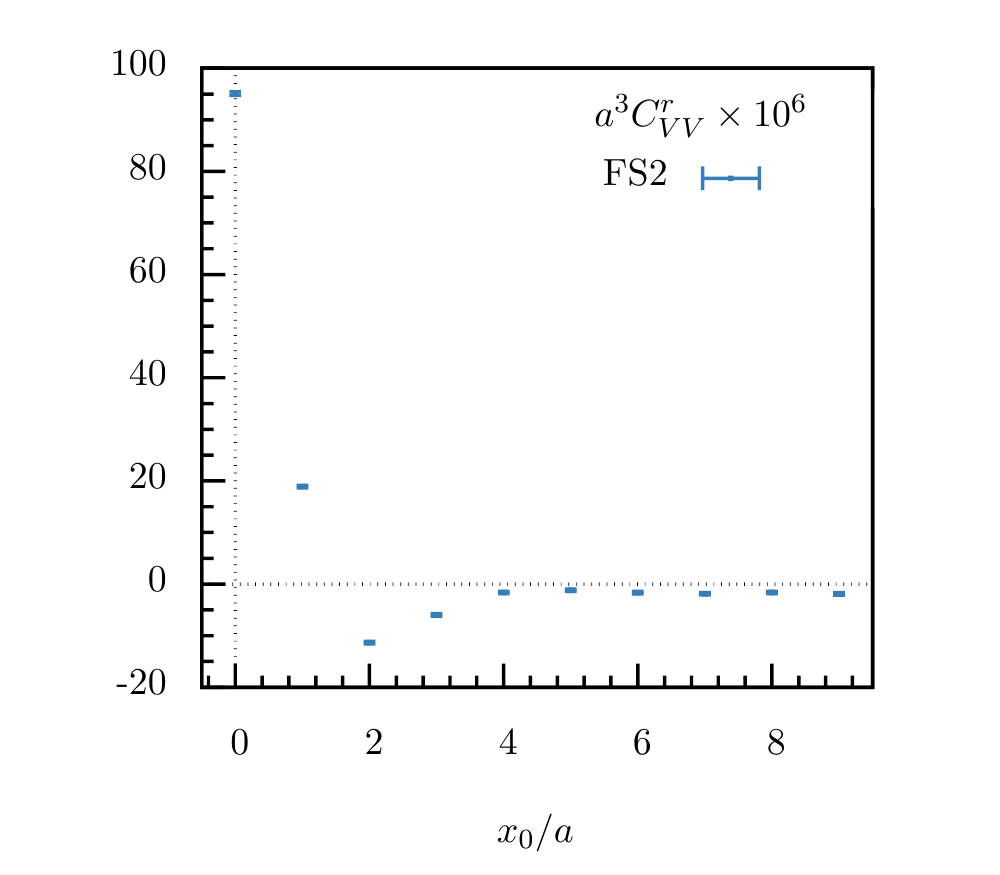}
\caption{Left: variance of the disconnected contribution in 
  Eq.~\eqref{eq:twopt_iso} with $x_0/a=10$ using the standard estimator (red filled
  squares) versus the number of sources, and the FS2 estimator (blue open squares)
  versus the number of its evaluations per gauge configuration.
  Right: the disconnected contribution using the FS2 estimator from $N_\mathrm{cfg}=1200$
  gauge configurations. With the same number of configurations and
  the same numerical cost, no signal is observed with the standard estimator.}
\label{Fig:twopt_iso} 
\end{center}
\end{figure}
In spectroscopic applications, disconnected diagrams arise generically in
isoscalar channels. The vector channel, for instance, contains the
contribution
\begin{equation}
  \label{eq:twopt_iso}
  C^{r}_{VV}(x_0) = - \frac{L^3}{3L_0}\sum_{k=1}^{3}
    \sum_{y_0} a\,
    \langle\bar t_{\gamma_k,r}(x_0+y_0)\, \bar t_{\gamma_k,r}(y_0)\rangle.
\end{equation}
To evaluate this correlation function, we use the FS2 estimator introduced in
section~\ref{sec:FSEs} for both single-propagator traces. In the left plot of 
Fig.~\ref{Fig:twopt_iso} we show the variances of the standard estimator
(filled symbols) against the number of sources, and the improved FS2 estimator
(open symbols) against the number of its evaluations per gauge configuration.
The gauge variance is approached with about
$N_s\sim256$ evaluations of the FS2 estimator, similarly to the
case of the one-point function of section~\ref{sec:FSEs}.
In this case, while the disconnected piece gives only a small contribution to
the isoscalar channel at intermediate hadronic distances, its variance quickly dominates
the statistical error at  large distances. The improved estimator thus allows the full
correlation function to be resolved at much larger distances.

\section{Conclusions}
The numerical computation of disconnected Wick contractions is challenging in lattice QCD
because (a) their variances are dominated by the vacuum contribution, which in turn implies
that statistical errors remain constant with the distance of the disconnected pieces while the
signal typically decreases exponentially, and (b) averaging each disconnected
sub-diagram over the volume tends to be numerically expensive because the quark propagators must
be re-computed at each lattice point.

A milestone for solving the second problem was the introduction of random-noise
estimators~\cite{Bitar:1988bb,Dong:1993pk,Michael:1998sg} which allow one to sum over many
or all source points stochastically. However for single-propagator traces, the simplest among the
disconnected sub-diagrams, such estimators tend to have variances which are typically orders of
magnitude larger than the intrinsic gauge noise. An a priori theoretical analysis of the variances
is thus mandatory for deciding how to define exactly the stochastic observables.   

Luckily the random-noise contribution to the variances can be
re-expressed in the form of simple integrated correlation functions of local composite operators, a fact which
allows us to use the quantum field theory machinery for analyzing the origin of the statistical errors
and eventually to reduce them.

As a result, we have introduced new stochastic observables for single-propagator traces: the split-even
and the frequency-splitting estimators for difference of two traces and for single traces respectively.
The former needs from a few random sources for the pseudoscalar density up to $O(100)$
for the vector current to approach the gauge noise. The reduction in numerical cost with respect to the
standard estimator ranges from one order of magnitude for the scalar and pseudoscalar densities up to around
two orders of magnitude or more for the axial and vector currents as well as for the tensor bilinear.
Just one or a few evaluations of the frequency-splitting estimators are needed for the variances of the
scalar, pseudoscalar and tensor bilinears to be comparable to the gauge noise, while for the axial-vector and
vector currents $O(10)$ and $O(100)$ evaluations are required to reach the same goal. In this case the
reduction of the computational cost with respect to the standard estimator is of one order of magnitude or
so depending on the bilinear. In all cases considered the variances of the stochastic estimators
reach the level of the intrinsic gauge noise with a moderate number of evaluations per gauge
configuration.

The use of these new estimators significantly speeds up the computation of disconnected fermion Wick contractions
which contribute to many physics processes at the forefront of research in particle and nuclear physics:
the hadronic contribution to the muon anomalous magnetic moment, $K\rightarrow\pi\pi$ decays, nucleon form factors,
quantum electrodynamics and strong isospin-breaking contributions to hadronic matrix elements, $\eta'$ propagator, etc.
As an example we have shown their potential for computing the disconnected contribution to 
the light-quark contribution to the muon anomalous magnetic moment and to the correlator of two
singlet vector currents. Theory suggests large gains for disconnected three and higher point correlation functions as well.
To solve or mitigate the problem (a) alluded
to at the beginning of this section, the next step is to combine these estimators with the newly proposed multi-level
integration in the presence of fermions~\cite{Ce:2016idq,Ce:2016ajy}.

\section{Acknowledgments}
Simulations have been performed on the PC clusters Marconi at CINECA
(CINECA-INFN and CINECA-Bicocca agreements) and Wilson at Milano-Bicocca.
We thank these institutions for the computer resources and the technical
support. We are grateful to our colleagues within the CLS initiative for
sharing the ensembles of gauge configurations with two dynamical flavours.
L.G. and T. H. acknowledge partial support by the INFN project
``High performance data network''.

\appendix

\section{$O(a)$-improved Wilson-Dirac operator\label{app:Dw}}
The massive $O(a)$-improved Wilson-Dirac operator is defined
as~\cite{Sheikholeslami:1985ij,Luscher:1996sc}
\be\label{eq:wdo}
D_m = D_\mathrm{w} + \delta D_\mathrm{v} + m\; ,
\ee
where $m$ is the bare quark mass, $D_\mathrm{w}$ is the massless Wilson-Dirac
operator
\begin{equation}
D_\mathrm{w}=
\frac{1}{2}\left\{ \gamma_\mu(\nabla^*_\mu+\nabla_\mu)- \nabla^*_\mu\nabla_\mu\right\}\; ,
\end{equation}
$\gamma_\mu$ are the Dirac matrices, and the summation over repeated indices is understood.
The covariant forward and backward derivatives 
$\nabla_\mu$ and $\nabla^*_\mu$ are defined to be
\be
a \nabla_{\mu}\psi(x) =  U_\mu(x)\psi(x+\hat{\mu})-\psi(x)\; ,\quad
a \nabla^*_\mu\psi(x) =  \psi(x)-  U_\mu^\dagger(x-\hat{\mu})
\psi(x-\hat{\mu})\; ,
\ee
where $U_\mu(x)$ are the link fields. The clover term is defined as 
\be
\delta D_\mathrm{v}\psi(x) =  a\, c_{_{\rm SW}} \frac{i}{4} \sigma_{\mu\nu} \widehat F_{\mu\nu}(x)\psi(x)\; ,\\[0.25cm]
\ee
where the field strength of the gauge field is
\be\label{eq:Fmunu}
a^2 \widehat F_{\mu\nu}(x) = \frac{1}{8} \{Q_{\mu\nu}(x) -  Q_{\nu\mu}(x)\} 
\ee
with
\bea
Q_{\mu\nu}(x) & = & 
U_\mu(x)\, U_\nu(x+ \hat\mu)\, U^\dagger_\mu(x + \hat\nu)\,U^\dagger_\nu(x)\nonumber\\[0.125cm]
& + & U_\nu(x)\, U^\dagger_\mu(x-\hat\mu+\hat\nu)\, U^\dagger_\nu(x - \hat\mu)\,
U_\mu(x-\hat\mu)\\[0.125cm]
& + & U^\dagger_\mu(x-\hat\mu)\, U^\dagger_\nu(x-\hat\mu-\hat\nu)\, 
U_\mu(x-\hat\mu-\hat\nu)\,U_\nu(x-\hat\nu)\nonumber\\[0.125cm]
& + & 
U^\dagger_\nu(x-\hat\nu)\, U_\mu(x-\hat\nu)\, 
U_\nu(x + \hat\mu - \hat\nu)\, U^\dagger_\mu(x)\; . \nonumber
\eea

\subsection{Hopping expansion\label{app:hpe}}
By applying the standard even-odd decomposition of the
Wilson--Dirac operator
\begin{equation}\label{eq:Deo_dec}
D_m=\left ( 
\begin{matrix} 
D_{\rm ee} & D_{\rm eo} \\
D_{\rm oe} & D_{\rm oo} \\
\end{matrix}
\right )\; , 
\end{equation}
see Ref.~\cite{Luscher:2010ae} for unexplained notation, 
it is straightforward to verify that
\be
D_m^{-1} = \frac{1}{D_{\rm ee} + D_{\rm oo}}\, \frac{1}{1-H_m}\; , 
\ee
where
\be
H_m = -\left[D_{\rm eo} D_{\rm oo}^{-1} + D_{\rm oe} D_{\rm ee}^{-1}  \right]\; ,  
\ee
and for clarity the subscript $m$ has been omitted in the block matrices of the
even/odd decomposition defined in Eq.~(\ref{eq:Deo_dec}). It follows that 
\be\label{eq:HPEdec}
D_m^{-1} = M_{2n,m} + D_m^{-1} H^{2n}_m\; ,   
\ee
where
\be\label{eq:HPEM}
M_{2n,m} = \frac{1}{D_{\rm ee} + D_{\rm oo}} \sum_{k=0}^{2n-1} H^k_m\; . 
\ee

\section{Bilinear chains in the free case\label{app:free}}
The propagator of a free Wilson fermion is 
\be
S(x-y) = \langle \psi(x) \bar\psi(y) \rangle =
\int_{\it BZ}\frac{d^4 p}{(2\pi)^4}\, K^{-1}(p)\, e^{i p (x-y)}\; ,  
\ee
where
\be
K(p) = i \gamma_\mu \bar p_\mu + M(p)\; , \qquad M(p) = m + \frac{a}{2} \hat p^2\; , 
\ee
with
\be
\bar p_\mu = \frac{1}{a}\sin(p_\mu a)\; , \quad
\hat p_\mu = \frac{2}{a}\sin\left(\frac{p_\mu a}{2}\right)\; , 
\ee
and as usual $\bar p^2 = \bar p_\mu \bar p_\mu$ and $\hat p^2 = \hat p_\mu \hat p_\mu$.

\subsection{Two-point correlators}
The integrated two-point correlation functions of non-singlet bilinears are 
\bea
\sum_{\bf x} a^3 \langle O_{_{\Gamma, rr'}} (0,{\bf x}) O_{_{\Gamma, r'r}} (0) \rangle & = &
-12 \int_{\it BZ} \frac{d^3 {\bf p}}{(2\pi)^3} \Big\{b_{_\Gamma}\, G({\bf p},m_r)\, G({\bf p},m_{r'}) 
+\nonumber\\
& & \hspace{2.75cm} c_{_\Gamma}\,  \bar{\bf p}^2 F({\bf p},m_r)\, F({\bf p},m_{r'})  \Big\}
\eea
where
\be
F({\bf p},m) =  \int_{\it BZ} \frac{d p_0}{2\pi} \frac{1}{\bar p^2 + M^2(p)}\, , \qquad
G({\bf p},m) =  \int_{\it BZ} \frac{d p_0}{2\pi} \frac{M(p)}{\bar p^2 + M^2(p)}
\ee
and 
\be
b_{_\Gamma} = \left\{\begin{array}{rl}
1 & \Gamma=I,\gamma_5,\gamma_\mu,\sigma_{\mu\nu}\\
-1 & \Gamma=\gamma_\mu\gamma_5\\
\end{array}\right.
\;\qquad
c_{_\Gamma} = \left\{\begin{array}{ll}
-1 & \Gamma=I\\
1 & \Gamma=\gamma_5\\
1 -\frac{2}{3}(1-\delta_{\mu 0}) & \Gamma=\gamma_\mu,\gamma_\mu\gamma_5\\
-1 +\frac{2}{3}(2-\delta_{\mu 0}-\delta_{\nu 0}) & \Gamma=\sigma_{\mu\nu}\\ 
\end{array}\right.
\ee
By using 
\be
 K^\dagger(p) K(p) = K(p) K^\dagger(p) =
\bar p^2 + M^2(p)\; , \qquad K^\dagger(p) = \gamma_5 K(p) \gamma_5\, , 
\ee
it is straightforward to obtain 
\be
\sum_y a^4 \langle P_{rr'} (y) P_{r'r} (0) \rangle = -12 \int_{\it BZ}
\frac{d^4 p}{(2\pi)^4} \frac{\bar p^2+M_r(p) M_{r'}(p)}{[\bar p^2 + M^2_r(p)][\bar p^2 + M^2_{r'}(p)]}\nonumber\; ,
\ee
where $M_r(p)$ is evaluated at the bare mass of $m_r$.

\subsection{Four-point correlators}
The integrated four-point correlation functions of non-singlet bilinears we
are interested in are
\bea
& & \sum_{y_1,{\bf y_2},y_3} a^{11} \big\langle  P_{rr'}(y_1)\, O_{_{\Gamma, r's'}}(0,{\bf y_2})\,
P_{s's}(y_3) \, O_{_{\Gamma, sr}} (0) \big\rangle =\nonumber\\[0.25cm]
& & -12\, a_\Gamma^2 \int_{\it BZ} \frac{d^3 {\bf p}}{(2\pi)^3} 
F({\bf p},m_r) F({\bf p},m_s) 
\label{eq:RNV7free}
\eea
and 
\bea
& & \sum_{y_1,{\bf y_2},y_3} a^{11}
\big\langle  S_{rs}(y_1)\, O_{_{\Gamma, ss'}}(0,{\bf y_2})\,
S_{s'r'}(y_3) \, O_{_{\Gamma, r'r}} (0) \big\rangle =\nonumber\\[0.25cm]
& & -12 \int_{\it BZ} \frac{d^3 {\bf p}}{(2\pi)^3}
\Big\{b_{_\Gamma}\, [K({\bf p},m_r,m_s)]^2\, +
c_{_\Gamma}\,  \bar{\bf p}^2 [H({\bf p},m_r,m_s)]^2 \Big\}
\eea
where
\bea
H({\bf p},m_r,m_s) & = &  \int_{\it BZ} \frac{d p_0}{2\pi}
\frac{M_r(p)+M_s(p)}{[\bar p^2 + M^2_r(p)][\bar p^2 + M^2_s(p)]}\; ,\\[0.25cm]
K({\bf p},m_r,m_s) & = & \int_{\it BZ} \frac{d p_0}{2\pi}
\frac{M_r(p) M_s(p) - \bar p^2}{[\bar p^2 + M^2_r(p)][\bar p^2 + M^2_s(p)]}\; ,
\eea
and $m_{r'}=m_r$ and $m_{s'}=m_s$. Finally
\bea
& & \sum_{y_1,y_2,y_3} a^{12} \langle S_{rs}(y_1) P_{ss'} (y_2) S_{s'r'}(y_3) P_{r'r} (0) \rangle =
\label{eq:freePSPS}\\[0.25cm]
& & -12 \int_{\it BZ}
\frac{d^4 p}{(2\pi)^4} \frac{1}{\{\bar p^2 + M_r^2(p)\}\{\bar  p^2 + M^2_s(p)\}}\nonumber
\eea
where again we are interested in the case $m_{r'}=m_r$ and $m_{s'}=m_s$.

\subsection{Variance of the HPE remainder}
In the free theory, the variance of the noisy estimator of the remainder in
Eq.~(\ref{eq:tauhR}) is
\bea
& & \sigma^2_{\bar \tau^{R}_{_{\Gamma,m}}} = \frac{6}{L^3 N_s} \int \frac{d^3 p}{(2\pi)^3}
\Big\{J_{0,n}({\bf p},m) J_{1,n}({\bf p},m) + [I_{1,2n}({\bf p},m)]^2 a_\Gamma^2 b_\Gamma +\nonumber\\[0.25cm]
& &\hspace{4.2cm}  [I_{0,2n}({\bf p},m)]^2\, \bar{\bf p}^2\, a_\Gamma^2\, c_\Gamma \Big\}
\label{eq:sigmaRFR}
\eea
where
\bea
J_{0,n}({\bf p},m) & = &  \int_{\it BZ} \frac{d p_0}{2\pi}\;
\Big\{(c_{1,n}(p))^2 \bar p^2 + (c_{0,n}(p))^2\Big\}\\[0.375cm]
J_{1,n}({\bf p},m) & = &  \int_{\it BZ} \frac{d p_0}{2\pi}\;
\frac{(c_{1,n}(p))^2 \bar p^2 + (c_{0,n}(p))^2}{\bar p^2 + M^2(p)}\\[0.375cm]
I_{0,n}({\bf p},m) & = &  \int_{\it BZ} \frac{d p_0}{2\pi}\;
\frac{c_{1,n}(p) M(p) - c_{0,n}(p)}{\bar p^2 + M^2(p)}\\[0.375cm]
I_{1,n}({\bf p},m) & = &  \int_{\it BZ} \frac{d p_0}{2\pi}\;
\frac{c_{1,n}(p) \bar p^2  + c_{0,n}(p) M(p) }{\bar p^2 + M^2(p)}
\eea
with 
\be\label{eq:cmn}
c_{m,n} = \sum_{k=0}^{(n-m)/2} \binom{n}{2k+m} (c_{0,1})^{n-(2k+m)} (c_{1,1})^{2k+m} (- {\bar p}^2)^{k}\;, \quad m=0,1\; ,
\ee
and
\be
c_{0,1}(p) = \frac{4 - a^2 \hat p^2/2}{a m+4}\; , \qquad
c_{1,1}(p) = - \frac{a}{a m+4}\;, 
\ee
where the highest $k$ in the sum (\ref{eq:cmn}) is the last integer value,
i.e. either $(n-m)/2$ or $(n-m-1)/2$. If the noise estimator of the remainder in
Eq.~(\ref{eq:tauhR}) would have been
defined by applying $H^{2n}_m$ to one source vector only, its variance would be
as in Eq.~(\ref{eq:sigmaRFR}) but with the replacement
$J_{0,n}({\bf p},m) J_{1,n}({\bf p},m)\rightarrow (1/a) J_{1,2n}({\bf p},m)$.

\section{Exact computation of the first $2n$ terms in the HPE\label{app:mpls}}
The matrix $M_{2n,m}$ in Eq.~(\ref{eq:HPEmain}) is sparse. Its diagonal elements can thus
be computed with a few applications of $M_{2n,m}$ on a well chosen set of probing
vectors. Following Ref.~\cite{Tang:2010}, if for a matrix ${\cal M}$ there
exist $K$ probing vectors $v^0,\ldots,v^{K-1}$ which
satisfy\footnote{In Ref.~\cite{Stathopoulos:2013aci} probing vectors were introduced
in the context of lattice QCD to define stochastic estimators of traces of the full quark
propagator.}
\begin{align}
    \sum_{k=0}^{K-1} v^k_iv^k_j &= \delta_{ij}\qquad\textrm{for
    all}\quad i,j\quad\textrm{where}\quad {\cal M}_{ij}\neq0,
    \label{eq:probing_orthog}
\end{align}
then the diagonal elements of ${\cal M}$ are given by (no summation over $i$)
\begin{align}
    {\cal M}_{ii} &= \sum_{k=0}^{K-1} v_i^ku^k_i, \qquad
    \textrm{where} \qquad u^k = {\cal M} v^k.
    \label{eq:probing}
\end{align}
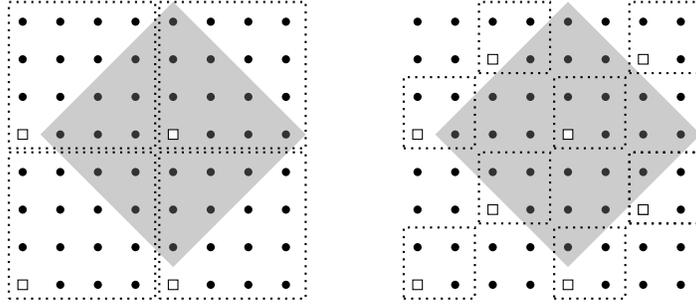
\begin{figure}[t]
    \centering
    \begin{tikzpicture}[scale=0.5]
      \tikzset{dstyle/.style={shape=circle,fill=black,scale=0.3}}
      \tikzset{ostyle/.style={shape=rectangle,draw,fill=white,scale=0.5}}
      \foreach \x in {0,...,7}
      \foreach \y in {0,...,7}
      {
        \node[dstyle] (\x-\y) at (\x,\y) {};
      }
      \node[diamond,fill=gray,opacity=0.4,scale=7] at (4,4) {};
      \foreach \x/\y in {0/0,4/4,0/4,4/0}
      {
        \node[ostyle] at (\x,\y) {};
        \node (p\x\y) at (\x+3,\y+3) {};
        \node[draw,thick,dotted,fit=(\x-\y) (p\x\y)] {};
      }
    \end{tikzpicture}%
    \hspace{3em}
    \begin{tikzpicture}[scale=0.5]
      \tikzset{dstyle/.style={shape=circle,fill=black,scale=0.3}}
      \tikzset{ostyle/.style={shape=rectangle,draw,fill=white,scale=0.5}}
      \tikzset{oostyle/.style={shape=rectangle,draw,fill,scale=0.5}}
      \foreach \x in {0,...,7}
      \foreach \y in {0,...,7}
      {
        \node[dstyle] (\x-\y) at (\x,\y) {};
      }
      \node[diamond,fill=gray,opacity=0.4,scale=7] at (4,4) {};
      \foreach \x/\y in {0/0,4/4,0/4,4/0,2/2,6/2,6/6,6/2,2/6}
      {
        
        \node[ostyle] at (\x,\y) {};
        \node (p\x\y) at (\x+1,\y+1) {};
        \node[draw,thick,dotted,fit=(\x-\y) (p\x\y)] {};
      }
    \end{tikzpicture}
    \caption{Two schemes of probing vectors suitable for a matrix whose generic column
      has non-zero entries on the sites within the shaded region. Left: the probing vectors
      have support, e.g. white squares, on one point of each of the squared blocks. The
      column of the matrix $M_{2n,m}$ corresponding to $(x/a)=(4,4)$ has non-zero entries only
      in the shaded region where the probing vector is zero. Right: as on the left but for
      smaller blocks labeled even and odd.
    }
    \label{fig:hopping}
\end{figure}
The non-zero elements of $M_{2n,m}$ are those that connect two lattice sites
$x$ and $y$ with $\norm{x-y}_1<na$, while the matrix is dense in the spin and colour indices. For a lattice
which can be decomposed in hypercubic blocks of size $(2na)^4$, an obvious scheme to define the set
of probing vectors which satisfies the condition
(\ref{eq:probing_orthog}) is
\begin{align}
    v^k(x) &=
    \begin{cases}
        1 \qquad k=i_{cs}+ 12\, l_{2n}(x) \\
        0 \qquad\textrm{otherwise}
    \end{cases}
    \label{eq:spatial_dilution}
\end{align}
where $i_{cs}=1,\dots,12$ indicates the spin-colour index and $l_{2n}(x)=(x_0/a)\bmod 2n+2n\cdot [(x_1/a) \bmod 2n]+\ldots$
is the lexicographical index labeling the sites in any given block. This scheme,
illustrated in Fig.~\ref{fig:hopping} for $n=2$, requires
$K=192\, n^4$ probing vectors because one vector is required for each of the spin-colour
components for every site in the block.

A more efficient scheme, already outlined in Ref.~\cite{Giusti:2018vxm}, is depicted in the right-hand
panel of Fig.~\ref{fig:hopping}, where even-odd blocks of half the linear size of the previous
ones are introduced.
The probing vectors are defined by
\begin{align}
    v^k(x) &=
    \begin{cases}
      1 \qquad k=i_{cs}+ 12\,\{p+ 2\, l_{n}(x)\}\\
        0 \qquad\textrm{otherwise}
    \end{cases}
    \label{eq:eo_spatial_dilution}
\end{align}
where as before $i_{cs}$ indicates the spin-colour index, $p=0,1$ is the parity of the block, and 
again $l_{n}(x)$ is the lexicographical index labeling the sites in any given block.
This scheme requires just $K=24\, n^4$ vectors, which is a factor $8$ fewer than the first one.
\bibliographystyle{JHEP}
\bibliography{mb.bib}

\providecommand{\href}[2]{#2}\begingroup\raggedright\begin{thebibliography}{10}

\bibitem{Ce:2016idq}
M.~{C\`e}, L.~Giusti, and S.~Schaefer, {\it {Domain decomposition, multi-level
  integration and exponential noise reduction in lattice QCD}},  {\em Phys.
  Rev.} {\bf D93} (2016), no.~9 094507,
  [\href{http://arxiv.org/abs/1601.04587}{{\tt arXiv:1601.04587}}].

\bibitem{Ce:2016ajy}
M.~Cè, L.~Giusti, and S.~Schaefer, {\it {A local factorization of the fermion
  determinant in lattice QCD}},  {\em Phys. Rev.} {\bf D95} (2017), no.~3
  034503, [\href{http://arxiv.org/abs/1609.02419}{{\tt arXiv:1609.02419}}].

\bibitem{Bitar:1988bb}
K.~Bitar, A.~D. Kennedy, R.~Horsley, S.~Meyer, and P.~Rossi, {\it {The {QCD}
  Finite Temperature Transition and Hybrid Monte Carlo}},  {\em Nucl. Phys.}
  {\bf B313} (1989) 348--376.

\bibitem{Dong:1993pk}
S.-J. Dong and K.-F. Liu, {\it {Stochastic estimation with Z(2) noise}},  {\em
  Phys. Lett.} {\bf B328} (1994) 130--136,
  [\href{http://arxiv.org/abs/hep-lat/9308015}{{\tt hep-lat/9308015}}].

\bibitem{Michael:1998sg}
{\bf UKQCD} Collaboration, C.~Michael and J.~Peisa, {\it {Maximal variance
  reduction for stochastic propagators with applications to the static quark
  spectrum}},  {\em Phys. Rev.} {\bf D58} (1998) 034506,
  [\href{http://arxiv.org/abs/hep-lat/9802015}{{\tt hep-lat/9802015}}].

\bibitem{Luscher:2004fu}
M.~L{\"u}scher, {\it {Topological effects in QCD and the problem of short
  distance singularities}},  {\em Phys. Lett.} {\bf B593} (2004) 296--301,
  [\href{http://arxiv.org/abs/hep-th/0404034}{{\tt hep-th/0404034}}].

\bibitem{DelDebbio:2006cn}
L.~Del~Debbio, L.~Giusti, M.~Luscher, R.~Petronzio, and N.~Tantalo, {\it {QCD
  with light Wilson quarks on fine lattices (I): First experiences and physics
  results}},  {\em JHEP} {\bf 02} (2007) 056,
  [\href{http://arxiv.org/abs/hep-lat/0610059}{{\tt hep-lat/0610059}}].

\bibitem{DelDebbio:2007pz}
L.~Del~Debbio, L.~Giusti, M.~Luscher, R.~Petronzio, and N.~Tantalo, {\it {QCD
  with light Wilson quarks on fine lattices. II. DD-HMC simulations and data
  analysis}},  {\em JHEP} {\bf 02} (2007) 082,
  [\href{http://arxiv.org/abs/hep-lat/0701009}{{\tt hep-lat/0701009}}].

\bibitem{Fritzsch:2012wq}
P.~Fritzsch, F.~Knechtli, B.~Leder, M.~Marinkovic, S.~Schaefer, R.~Sommer, and
  F.~Virotta, {\it {The strange quark mass and Lambda parameter of two flavor
  QCD}},  {\em Nucl. Phys.} {\bf B865} (2012) 397--429,
  [\href{http://arxiv.org/abs/1205.5380}{{\tt arXiv:1205.5380}}].

\bibitem{Engel:2014eea}
G.~P. Engel, L.~Giusti, S.~Lottini, and R.~Sommer, {\it {Spectral density of
  the Dirac operator in two-flavor QCD}},  {\em Phys. Rev.} {\bf D91} (2015),
  no.~5 054505, [\href{http://arxiv.org/abs/1411.6386}{{\tt arXiv:1411.6386}}].

\bibitem{Thron:1997iy}
C.~Thron, S.~J. Dong, K.~F. Liu, and H.~P. Ying, {\it {Pade - Z(2) estimator of
  determinants}},  {\em Phys. Rev.} {\bf D57} (1998) 1642--1653,
  [\href{http://arxiv.org/abs/hep-lat/9707001}{{\tt hep-lat/9707001}}].

\bibitem{McNeile:2000xx}
{\bf UKQCD} Collaboration, C.~McNeile and C.~Michael, {\it {Mixing of scalar
  glueballs and flavor singlet scalar mesons}},  {\em Phys. Rev.} {\bf D63}
  (2001) 114503, [\href{http://arxiv.org/abs/hep-lat/0010019}{{\tt
  hep-lat/0010019}}].

\bibitem{Bali:2009hu}
G.~S. Bali, S.~Collins, and A.~Schafer, {\it {Effective noise reduction
  techniques for disconnected loops in Lattice QCD}},  {\em Comput. Phys.
  Commun.} {\bf 181} (2010) 1570--1583,
  [\href{http://arxiv.org/abs/0910.3970}{{\tt arXiv:0910.3970}}].

\bibitem{Giusti:2008vb}
L.~Giusti and M.~L{\"u}scher, {\it {Chiral symmetry breaking and the
  Banks-Casher relation in lattice QCD with Wilson quarks}},  {\em JHEP} {\bf
  03} (2009) 013, [\href{http://arxiv.org/abs/0812.3638}{{\tt
  arXiv:0812.3638}}].

\bibitem{DellaMorte:2017dyu}
M.~Della~Morte, A.~Francis, V.~Gülpers, G.~Herdoíza, G.~von Hippel, H.~Horch,
  B.~Jäger, H.~B. Meyer, A.~Nyffeler, and H.~Wittig, {\it {The hadronic vacuum
  polarization contribution to the muon $g-2$ from lattice QCD}},  {\em JHEP}
  {\bf 10} (2017) 020, [\href{http://arxiv.org/abs/1705.01775}{{\tt
  arXiv:1705.01775}}].

\bibitem{Boucaud:2008xu}
{\bf ETM} Collaboration, P.~Boucaud et~al., {\it {Dynamical Twisted Mass
  Fermions with Light Quarks: Simulation and Analysis Details}},  {\em Comput.
  Phys. Commun.} {\bf 179} (2008) 695--715,
  [\href{http://arxiv.org/abs/0803.0224}{{\tt arXiv:0803.0224}}].

\bibitem{Dinter:2012tt}
{\bf ETM} Collaboration, S.~Dinter, V.~Drach, R.~Frezzotti, G.~Herdoiza,
  K.~Jansen, and G.~Rossi, {\it {Sigma terms and strangeness content of the
  nucleon with $N_f=2+1+1$ twisted mass fermions}},  {\em JHEP} {\bf 08} (2012)
  037, [\href{http://arxiv.org/abs/1202.1480}{{\tt arXiv:1202.1480}}].

\bibitem{openQCD1.6}
 \href{http://arxiv.org/abs/{http://luscher.web.cern.ch/luscher/openQCD/}}{{\tt
  {http://luscher.web.cern.ch/luscher/openQCD/}}}.

\bibitem{Giusti:2004yp}
L.~Giusti, P.~Hernandez, M.~Laine, P.~Weisz, and H.~Wittig, {\it {Low-energy
  couplings of QCD from current correlators near the chiral limit}},  {\em
  JHEP} {\bf 04} (2004) 013, [\href{http://arxiv.org/abs/hep-lat/0402002}{{\tt
  hep-lat/0402002}}].

\bibitem{DeGrand:2004qw}
T.~A. DeGrand and S.~Schaefer, {\it {Improving meson two point functions in
  lattice QCD}},  {\em Comput. Phys. Commun.} {\bf 159} (2004) 185--191,
  [\href{http://arxiv.org/abs/hep-lat/0401011}{{\tt hep-lat/0401011}}].

\bibitem{Blum:2012uh}
T.~Blum, T.~Izubuchi, and E.~Shintani, {\it {New class of variance-reduction
  techniques using lattice symmetries}},  {\em Phys. Rev.} {\bf D88} (2013),
  no.~9 094503, [\href{http://arxiv.org/abs/1208.4349}{{\tt arXiv:1208.4349}}].

\bibitem{Bernecker:2011gh}
D.~Bernecker and H.~B. Meyer, {\it {Vector Correlators in Lattice QCD: Methods
  and applications}},  {\em Eur. Phys. J.} {\bf A47} (2011) 148,
  [\href{http://arxiv.org/abs/1107.4388}{{\tt arXiv:1107.4388}}].

\bibitem{DallaBrida:2018tpn}
M.~Dalla~Brida, T.~Korzec, S.~Sint, and P.~Vilaseca, {\it {High precision
  renormalization of the flavour non-singlet Noether currents in lattice QCD
  with Wilson quarks}},  {\em Eur. Phys. J.} {\bf C79} (2019), no.~1 23,
  [\href{http://arxiv.org/abs/1808.09236}{{\tt arXiv:1808.09236}}].

\bibitem{Gerardin:2018kpy}
A.~Gerardin, T.~Harris, and H.~B. Meyer, {\it {Nonperturbative renormalization
  and $O(a)$-improvement of the nonsinglet vector current with $N_f=2+1$ Wilson
  fermions and tree-level Symanzik improved gauge action}},  {\em Phys. Rev.}
  {\bf D99} (2019), no.~1 014519, [\href{http://arxiv.org/abs/1811.08209}{{\tt
  arXiv:1811.08209}}].

\bibitem{Bhattacharya:2005rb}
T.~Bhattacharya, R.~Gupta, W.~Lee, S.~R. Sharpe, and J.~M.~S. Wu, {\it
  {Improved bilinears in lattice QCD with non-degenerate quarks}},  {\em Phys.
  Rev.} {\bf D73} (2006) 034504,
  [\href{http://arxiv.org/abs/hep-lat/0511014}{{\tt hep-lat/0511014}}].

\bibitem{Sheikholeslami:1985ij}
B.~Sheikholeslami and R.~Wohlert, {\it {Improved Continuum Limit Lattice Action
  for QCD with Wilson Fermions}},  {\em Nucl. Phys.} {\bf B259} (1985) 572.

\bibitem{Luscher:1996sc}
M.~{L\"uscher}, S.~Sint, R.~Sommer, and P.~Weisz, {\it {Chiral symmetry and
  O(a) improvement in lattice QCD}},  {\em Nucl. Phys.} {\bf B478} (1996)
  365--400, [\href{http://arxiv.org/abs/hep-lat/9605038}{{\tt
  hep-lat/9605038}}].

\bibitem{Luscher:2010ae}
M.~L{\"u}scher, {\it {Computational Strategies in Lattice QCD}},  in {\em
  {Modern perspectives in lattice QCD: Quantum field theory and high
  performance computing. Proceedings, International School, 93rd Session, Les
  Houches, France, August 3-28, 2009}}, pp.~331--399, 2010.
\newblock \href{http://arxiv.org/abs/1002.4232}{{\tt arXiv:1002.4232}}.

\bibitem{Tang:2010}
J.~M. Tang and Y.~Saad, {\it A probing method for computing the diagonal of a
  matrix inverse},  {\em Numerical Linear Algebra with Applications} {\bf 19}
  (2012), no.~3 485--501,
  [\href{http://arxiv.org/abs/https://onlinelibrary.wiley.com/doi/pdf/10.1002/nla.779}{{\tt
  https://onlinelibrary.wiley.com/doi/pdf/10.1002/nla.779}}].

\bibitem{Stathopoulos:2013aci}
A.~Stathopoulos, J.~Laeuchli, and K.~Orginos, {\it {Hierarchical probing for
  estimating the trace of the matrix inverse on toroidal lattices}},
  \href{http://arxiv.org/abs/1302.4018}{{\tt arXiv:1302.4018}}.

\bibitem{Giusti:2018vxm}
L.~Giusti, T.~Harris, A.~Nada, and S.~Schaefer, {\it {Multi-level integration
  for meson propagators}},  in {\em {36th International Symposium on Lattice
  Field Theory (Lattice 2018) East Lansing, MI, United States, July 22-28,
  2018}}, 2018.
\newblock \href{http://arxiv.org/abs/1812.01875}{{\tt arXiv:1812.01875}}.

\end{thebibliography}\endgroup

\end{document}